\documentclass[preprint,showpacs,preprintnumbers,amsmath,amssymb]{revtex4}

\usepackage{graphicx}
\usepackage{dcolumn}
\usepackage{bm}

\begin{document}

\preprint{}

\title{Hubs and Clusters in the 
Evolving United States Internal Migration Network}

\author{Paul B. Slater}%
\email{slater@kitp.ucsb.edu}
\affiliation{%
ISBER, University of California, Santa Barbara, CA 93106\\
}%
\date{\today}
\newpage
\newpage
\begin{abstract}
Most nations of the world periodically publish
$N \times N$ origin-destination tables, recording the number of people
who lived in geographic subdivision $i$ at time $t$ and $j$ at $t+1$.
We have developed and widely applied to such national tables and 
other 
analogous 
(weighted, directed) 
socioeconomic networks, 
a two-stage--double-standardization 
and (strong component) hierarchical clustering--procedure.
Previous applications of this methodology
and related analytical issues are discussed. 
Its use is illustrated in a large-scale study, employing recorded 
United States internal migration flows between the 
3,000+ county-level units of the nation for the 
periods 1965-1970 and 1995-2000. Prominent, 
important features--such as 
''cosmopolitan hubs'' 
and ``functional regions''--are extracted 
from master dendrograms. The extent to
which such characteristics have varied over the intervening thirty years
is evaluated.
\end{abstract}


\maketitle
\section{Introduction}
A. L. Barab{\' a}si, in his 
recent popular book, ``Linked'', asserts that the emergence 
of {\it hubs} in networks is a surprising phenomenon that is ``forbidden by both the Erd{\"o}s-R{\'e}nyi and Watts-Strogatz models" \cite[p. 63]{linked} 
\cite[Chap. 8]{siegfried}.
Here, we indicate--and apply anew to extensive 
U. S. intercounty migration data--an analytical framework introduced
in 1974 that the distinguished computer scientist 
R. C. Dubes, in a review of the 
compilation of multitudinous results  \cite{tree}, asserted 
``might very well be the most successful application 
of cluster analysis'' \cite[p. 142]{dubes}. 
This {\it two}-stage methodology has proved insightful
in revealing--in addition, to functional clusters--hub-like structures 
in networks of (weighted, directed) internodal flows. 
This approach, together with its many diverse socioeconomic
applications, was documented in a large number of 
(subject-matter and technical) journal articles
(among them \cite{japan,france,winchester,science,partial,IO,college,gentileschi,metron,SEAS,qq,spain,schwarz,romania,russia,hirst1,seps,india}), 
as well as in the research institute monographs \cite{tree}, 
\cite{county}, \cite{tree2}. It has also been the subject of various
comments, criticisms  and discussions 
\cite{masser,kim,hirst2,findlay,seps,holmes1,holmes2,holmes3,telephone,baumann,clark,boyd,pandit,hoover,discernment} (cf. \cite{takayuki,noronha,corvers}).

Though this procedure is applicable 
in a wide variety of social-science 
settings \cite{tree,dubes}, it has been primarily 
used, in a demographic context, to study the 
{\it internal migration} tables published at regular periodic intervals by most of the nations of the world. 
These tables can be thought of as $N \times N$ (square) matrices, the entries ($m_{ij}$) of which are the number of people who
lived in geographic subdivision $i$ at time $t$ and $j$ at 
time $t+1$. (Some tables--but not all--have diagonal entries, $m_{ii}$, which may represent either the number of people who did move within
area $i$, or simply those who lived in $i$ both 
at $t$ and $t+1$. It can 
sometimes be of interest to compare analyses with zero-  and
nonzero-diagonal entries \cite{county}. However, this aspect will not 
be of any immediate concern to us here.)
We will principally be considering the case below of U. S. migration
tables for the periods 1965-1970 \cite{county} and 1995-2000 based on 3,000+
county-level units.
\section{Two-Stage Methodology}
\subsection{First Step: Double-Standardization of Raw Flows} \label{first}
In the {\it first} step (iterative proportional fitting procedure [IPFP] 
\cite{fienberg}),
the rows and columns of the 
table of flows 
are alternately  (biproportionally \cite{bacharach}) scaled 
to sum to a fixed number (say 1). Under broad conditions--to be discussed 
below--convergence occurs to a ``doubly-stochastic" 
(bistochastic) 
table, with row and column sums all 
{\it simultaneously} equal to 1 \cite{mosteller,louck,CSBZ,unistochastic,
romney,wong}. 
The purpose of the scaling is to 
remove overall (marginal) effects of size, and focus on relative, 
interaction  effects. 
Nevertheless, the {\it cross-product ratios} 
({\it relative odds}), $\frac{m_{ij} m_{kl}}{m_{il} m_{kj}}$, 
measures of association, are left {\it invariant}. 
Additionally, the entries of the
doubly-stochastic table provide 
{\it maximum  entropy} estimates of the original
flows, given the row and column constraints \cite{eriksson,macgill}.

For large {\it sparse} 
flow tables, only the nonzero entries, together with their
row and column coordinates are needed. Row and column (biproportional) 
multipliers can be iteratively computed by sequentially accessing the nonzero
cells \cite{parlett}. If the table is ``critically sparse'', various convergence difficulties may occur. Nonzero entries that are ``unsupported''--that is, not part of a set of $N$ nonzero entries, no two in the same row and 
column-- may converge to zero and/or the 
biproportional multipliers may not converge \cite[p. 19]{tree} \cite{sinkhorn} \cite[p. 171]{mirsky}.
The ``first strongly polynomial-time algorithm for matrix scaling'' 
was reported in \cite{linial}.

The scaling was successfully implemented, in our largest analysis,  
with a
$3,140 \times 3,140$ 1965-70 intercounty migration table--having 94.5\% 
of its entries, zero--for
the United States \cite{county,partial}, as well as for a more aggregate
$510 \times 510$ table (with {\it State Economic Areas} as  the basic 
unit) for the US for the same period \cite{SEAS}. ({\it Smoothing} procedures
could be used to modify the zero-nonzero structure 
of a flow table, particularly 
if it is critically sparse \cite{simonoff,boundary}. If one takes the 
second power of 
a doubly-stochastic matrix [as we in fact 
do in Sec.~\ref{square}], one obtains another such 
matrix--of predicted {\it two}-stage movements--but 
smoother in character. One might also 
consider standardizing the {\it i}th row [column] sum 
to be proportional to the number of non-zero entries in the 
{\it i}th row [column]--although we found considerable numerical 
difficulties when attempting this for the 1995-2000 U. S. intercounty
migration table [Sec.~\ref{newer}]. Another procedure--in line with
the Google page-ranking [``teleporting random walk''] 
procedure \cite{brinpage,langville}, that has been much studied and 
emulated--is to take some convex combination of the doubly-stochastic
table and the $N \times N$ table all the off-diagonal entries of which
are equal to $\frac{1}{N-1}$.)

\begin{figure}
\includegraphics{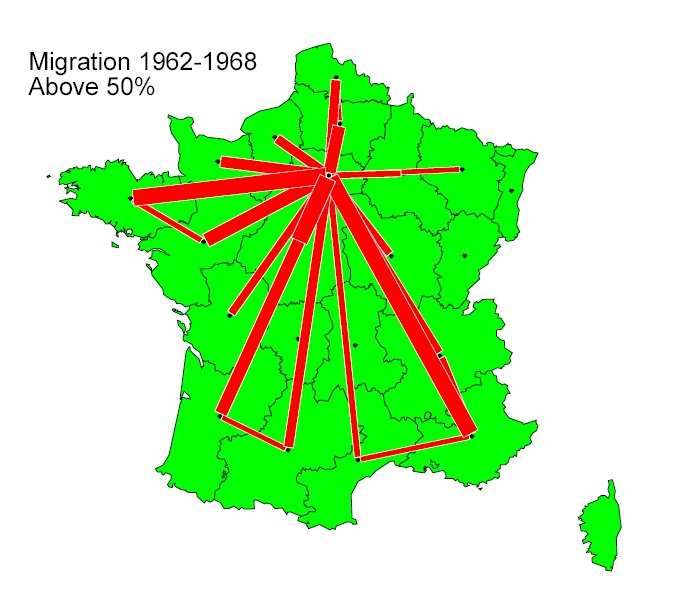}
\caption{\label{fig:waldofig1} Salient flows based on 
unadjusted 1962-68 French interprovincial migration table}
\end{figure}

\begin{figure}
\includegraphics{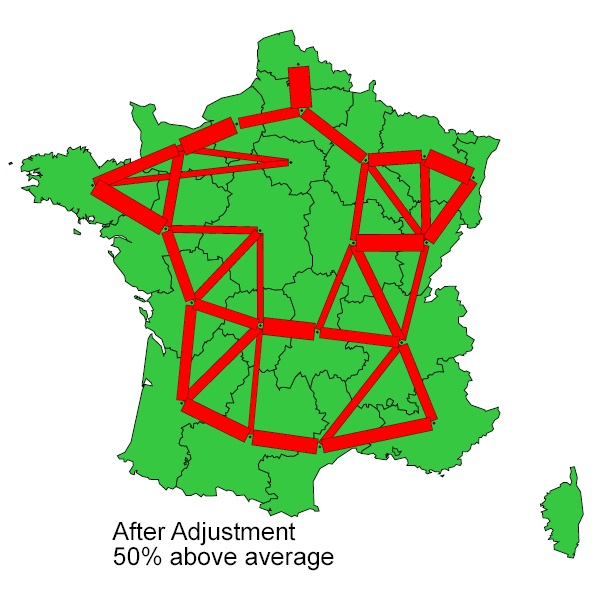}
\caption{\label{fig:waldofig2} Salient flows based on 
doubly-standardized 1962-68 French interprovincial migration table. Note the
increase in linkages between adjacent areas.}
\end{figure}

\begin{figure}
\includegraphics{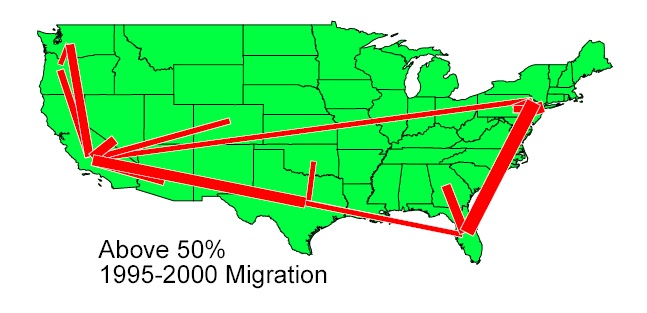}
\caption{\label{fig:waldofig3} Salient flows based on 
1965-70 unadjusted U. S. interstate migration table}
\end{figure}

\begin{figure}
\includegraphics{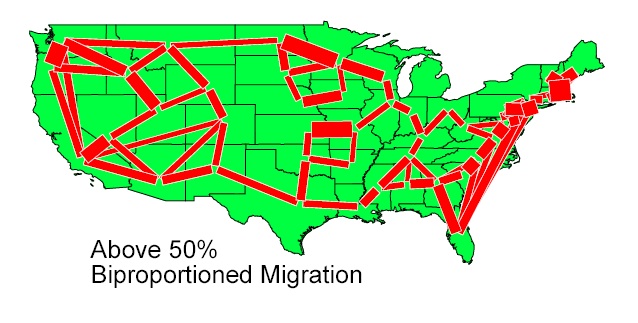}
\caption{\label{fig:waldofig4} Salient flows based on 1965-70 
doubly-standardized U. S. interstate migration table.
Note the
increase in linkages between adjacent areas.}
\end{figure}

Figs.~\ref{fig:waldofig1} and 
\ref{fig:waldofig2} give a graphic display of the 
effect of the double-standardization (biproportional adjustment) on 
a 1962-68 French interprovincial migration table 
(the island of Corsica is omitted). Additionally,
Figs.~\ref{fig:waldofig3} and \ref{fig:waldofig4} are comparable displays
for 1995-2000 U. S. interstate migration (Alaska, Hawaii and the 
District of Columbia are not included) (cf. \cite{maier}).
The process--employed by Waldo Tobler--to produce these four 
figures was the following: (1) the average
value ($A$) of all the entries 
in the (adjusted  or unadjusted) table was found; (2)
if the sum of the $ij$ and $ji$-entries of the corresponding table exceeded
$2 A$, a bar, the thickness of which is 
proportional to this sum is drawn connecting 
area $i$ to area $j$. We see
that in the two figures based on the double-standardization procedure,
linkages between adjacent areas are much more (suiting our purposes) 
strongly stressed than 
using the raw flows themselves.
\subsection{Second Step: Strong Component Hierarchical Clustering}
In the {\it second} step of the two-stage 
procedure, the doubly-stochastic matrix 
is converted to a series of {\it directed} 
(0,1) graphs (digraphs), by applying thresholds to its entries. 
As the thresholds   are 
progressively lowered, larger and larger {\it strong components} 
(a directed path existing from any member of a component 
to any other) of the resulting 
graphs are found. This process 
(a simple variant of well-known single-linkage [nearest-neighbor or min]
clustering \cite{gower1}) can be represented by the familiar dendrogram 
or tree diagram used in 
hierarchical cluster analysis and cladistics/phylogeny (cf. \cite{ozawa,hubert}). (The ``CLASSIC'' methodology proposed 
somewhat later by Ozawa--though couched in
rather different terminology--appears to be fully equivalent to ours. 
Ozawa found the procedure to be 
useful in ``the detection of gestalt clusters'' 
\cite{ozawa}.)
\subsection{Computer implementations}
A FORTRAN implementation of the two-stage process was given in 
\cite{leusmann}, as well as a realization 
in the SAS (Statistical Analysis System) 
framework \cite{chilko}. Subsequently, 
the noted computer scientist R. E. Tarjan 
\cite{schwartz} devised an $O(M (\log{N})^2)$
algorithm \cite{tarjan} for strong component hierarchical clustering, 
and, then, a further improved $O(M (\log{N}))$ 
method \cite{tarjan2}, 
where $N$ is the number of nodes and $M$ the number of edges of 
a directed graph. (These substantially improved upon 
the earlier works \cite{leusmann,chilko}, 
which 
required the 
computations of {\it transitive closures} of graphs--in terms of 
which the analysis of Ozawa \cite{ozawa} is phrased--and were 
$O(M N)$ in nature.) A FORTRAN coding--involving 
linked lists--of the improved Tarjan 
algorithm \cite{tarjan2} was presented
in \cite{tarjanslater}, and applied in the 
aforementioned 1965-70 US intercounty study \cite{county}. 
If the graph-theoretic (0,1)-structure of a network under study 
is {\it not} strongly connected 
\cite{hartfiel}, {\it independent} two-stage analyses of 
the subsystems of the network would be appropriate.

The {\it goodness-of-fit} of the dendrogram generated 
to the doubly-stochastic table itself
can be evaluated--and possibly employed, it would seem, as an optimization 
criterion (cf. \cite[p. 210]{hansen} \cite[Sec. 3]{cmn} \cite{radu}). 
Distances between nodes 
in the dendrogram satisfy the (stronger than {\it triangular}) 
{\it ultrametric} 
inequality, $d_{ij} \leq \max{(d_{ik},d_{jk})}$ \cite[p. 245]{johnson} 
\cite[eq. (2.2)]{rammal}.
We will examine issues pertaining to ultrametric fit and residuals
from such fits in Sec.~\ref{most}. (Costa and colleagues have studied
hierarchical aspects of ``complex networks'' \cite{costa,costa2}.)
\section{Empirical Results}
\subsection{{\it Cosmopolitan} or Hub-Like Units}
\subsubsection{Internal migration flows}
Geographic subdivisions (or groups of subdivisions) that enter into the 
bulk of the dendrograms produced 
by the two-stage procedure at the {\it weakest} levels are those with the 
{\it broadest} ties.
These are ``cosmopolitan", hub-like areas, 
a prototypical example being 
the French capital, Paris \cite[Sec. 4.1]{tree} \cite{france}. 
Similarly, 
in parallel analyses of other 
internal migration tables, the cosmopolitan/non-provincial natures
of London \cite{siegen}, 
Barcelona \cite{spain} \cite[Sec. 6.2, Figs. 36, 37]{tree}, 
Milan \cite{gentileschi} \cite[Sec. 6.3, Figs. 39, 40]{tree} 
(cf. \cite{metron}), Amsterdam 
\cite[p. 78]{tree} \cite{masser},
West Berlin \cite[p. 80]{tree}, Moscow (the city and the oblast as a unit) 
\cite{russia} 
\cite[Sec. 5.1 and Figs. 6, 7]{tree}, Manila (coupled with suburban Rizal) 
\cite{manila}, 
Bucharest \cite{romania}, 
{\^I}le-de-Montr{\'e}al \cite[p. 87]{tree}, 
Z{\"u}rich, Santiago, Tunis and Istanbul \cite{turkish} 
were--among
others--highlighted in the respective dendrograms for their nations
 \cite[Sec. 8.2]{tree} 
\cite[pp. 181-182]{qq} \cite[p. 55]{science}. In the 
1965-70 intercounty analysis
for the US, the most cosmopolitan entities were: (1) the 
{\it centrally}-located paired
Illinois counties of Cook (Chicago) and neighboring, suburban DuPage; 
(2) the nation's capital, Washington, D. C.; and (3) the paired South
Florida (retirement) counties of Dade (Miami) and Broward (Ft. Lauderdale) 
\cite{county,partial,fields}. In general, counties with large military 
installations, large college populations or state capitals 
also interacted broadly with other areas \cite[p. 153]{county}. 
Application of the two-stage methodology to 1965-66 London inter-borough
migration \cite{masser} indicated that the three inner boroughs of Kensington
and Chelsea, Westminster, and Hammersmith acted--as a unit--in a 
cosmopolitan manner \cite[Sec. 5.2, Fig. 10]{tree}. 
(In Sec. 8.2 and Table 16 of the anthology of results \cite{tree}, 
additional geographic units and groups of
units found to be cosmopolitan with regard to migration, are enumerated.)

It should be emphasized that 
although the indicated cosmopolitan areas may generally have 
relatively large populations, 
this can not, in and of itself, 
explain the wide national ties observed, since the 
double-standardization, in effect, renders all areas of equal overall size.
(However, to the extent that larger areas do have fewer zero entries in their
corresponding rows and columns, a bias to cosmpolitanism may 
in fact be present, 
which should be carefully considered. Possible corrections for bias were 
discussed above in Sec.~\ref{first}.)
If one were to obtain a (zero-diagonal) 
doubly-stochastic matrix, all the entries of which
were simply $\frac{1}{N-1}$, it would indicate complete 
{\it indifference} among migrants
as to where they come from and to where they go.
A maximally cosmopolitan unit would be one for which all the corresponding
row and column entries were $\frac{1}{N-1}$ (if all the diagonal
entries, $m_{ii}$, are {\it a priori} zero).
(It seems interesting to note that cosmopolitan areas appear to have
a certain {\it minimax} character, that is, the maximum doubly-stochastic
entry for the corresponding row and column tends to be minimized.)
\subsubsection{Trade and interindustry flows}
The nation of Italy possessed the broadest ties in a two-stage analysis
of the value of 1974 trade between 113 nations, followed by a closely-bound
group composed of the four Scandinavian countries \cite{schwarz} 
\cite[Sec. 5.6, Fig. 22]{tree}.
In a two-stage 
study (but using {\it weak} rather than strong components of the 
associated digraphs) of the 
1967 U. S. interindustry transaction table, the industry
with the broadest (most diffuse) ties was found to be Other Fabricated
Metal Products \cite{IO,gosling} \cite[pp. 13-18]{tree2}.
\subsubsection{Journal citations}
 In a two-stage analysis of 
22 mathematical journals, the {\it Annals of Mathematics} and {\it 
Inventiones Mathematicae} were strongly paired, while the {\it Proceedings of
the American Mathematical Society} was found to 
possess the broadest, most diffuse ties 
\cite{science}.

In a recent, large-scale ($N>6000$) journal-to-journal 
 citation analysis, decomposing ``the network into modules by compressing 
a description of the probability flow'', Rosvall and Bergstrom 
preliminarily {\it omitted } from their analysis 
the prominent journals {\it Science}, 
{\it Nature} and the 
{\it Proceedings of the National Academy of Sciences} 
 \cite[p. 1123]{rosvall}.
(Those are precisely the ones that would be expected to be ``cosmopolitan'' 
or hub-like in 
character, and to be highlighted in a corresponding
two-stage analysis.) Their rationale for 
the omission was that ``the broad scope of these journals otherwise creates an 
illusion of tighter connections among disciplines, when in fact few readers 
of the physics articles in {\it Science} also are close readers of the 
biomedical articles therein''. (In \cite[pp. 125-153]{tree2}, we reported
the results of a {\it partial} hierarchical clustering--not 
a two-stage analysis, but one originally designed and conducted 
by Henry G. Small and William 
Shaw--of citations between more than 3,000 journals. The clusters 
obtained there were compared with the actual 
subject matter classification employed
by the Institute for Scientific Information.)
\subsection{Functional Clusters of Units}
\subsubsection{Internal migration regions}
Geographically isolated (insular) areas--such as the Japanese islands of 
Kyushu and Shikoku \cite{japan}--emerged 
as well-defined {\it clusters} (regions)
of their constituent (seven and four, respectively) 
subdivisions (``prefectures'' in the Japanese case) 
in the dendrograms for the two-stage analyses, and similarly
 the Italian islands of Sicily and Sardinia 
\cite{gentileschi}, the North and South Islands of New Zealand, and the
Canadian islands of 
Newfoundland and Prince Edward Island \cite[p. 90]{tree}  
(cf. \cite{e,multiterminal}).
The eight counties of Connecticut, and other New England groupings, as  
further examples,  to be elaborated upon below, were
also very prominent in the highly disaggregated U. S. analysis \cite{county}. 
Relatedly, in a study based solely upon 
the 1968 movement of {\it college students} among
the fifty states, the six New England states were strongly clustered 
\cite[Fig. 1]{college}. Employing a 1963 Spanish interprovincial migration
table, well-defined regions were formed by the two provinces of
the Canary Islands, and the four provinces of Galicia \cite{spain} 
\cite[Sec. 6.2.1, Fig. 37]{tree}. 
The southernmost Indian states of Kerala and Madras (now Tamil Nadu) 
were strongly paired on the basis of 1961 interstate flows \cite{india}.
A detailed comparison between functional migration regions found by 
the two-stage procedure and those actually 
employed for administrative, political 
purposes in the corresponding nations is given in Sec. 8.1 and Table 15 of 
\cite{tree}. 

It should be noted that it is rare  that the two-stage
methodology yields a migration region 
composed of two or more noncontiguous subregions--even though no contiguity
information, of course,  is explicitly present in the flow table 
nor provided to the algorithm (cf. \cite{loglinear,boundary}).
A notable exception to this rule was the uniting of the northern 
Italian region of
Piemonte--the location of industrial Turin, where Fiat is based--with 
(poor) southern regions, {\it before} joining with central regions, in an 
aggregate 18-region  1955-70 study \cite{metron} 
\cite[p. 75]{tree} (cf. \cite{gentileschi}).
\subsubsection{Intermarriage and interindustry clusters}
 In a two-stage analysis of a $32 \times 32$ table of 
birthplace of bridegroom versus birthplace of bride of 1947 Australian
intermarriages \cite{price}, Greece and Cyprus were the strongest dyad 
\cite[Sec. 5.7, Fig. 25]{tree}.

In the 1967 US interindustry 
two-stage ({\it weak} component) analysis, two particularly salient pairs 
of functionally-linked industries were: (1)
Stone and Clay Products, and Stone and Clay Mining and Quarrying; and (2) 
Household Appliances and Service Industry Machines (the latter industry
purchases laundry equipment, refrigerators and freezers from the former)
\cite{IO,gosling} \cite[pp. 13-18]{tree2}. 

\section{Statistical Aspects}
It would be of interest to develop a theory--making use of 
the rich mathematical structure of doubly-stochastic 
matrices--by which the {\it statistical
significance} of apparent hubs and clusters 
in dendrograms produced by the two-stage procedure 
could be evaluated \cite[pp. 7-8]{county} \cite{bock}. 
In the geographic context of internal migration tables, where nearby areas 
have a strong distance-adversion predilection for binding, it seems unlikely 
that most clustering
results generated could be considered to be--in any standard 
sense--``random'' in nature. 
On the other hand, other types of ``origin-destination'' 
tables, such as those for 
{\it occupational} mobility \cite{duncan}, journal citations 
\cite{science} \cite[pp. 125-153]{tree2}, interindustry (input-output) flows 
\cite{IO} \cite{gosling}, brand-switches \cite[Sec. 9.6]{tree} \cite{rao}, 
crime-switches \cite[Sec. 9.7]{tree} \cite[Table XII]{blumstein}, and (Morse code) 
confusions \cite[Sec. 9.8]{tree} \cite{rothkopf}, among others, clearly lack such a geographic dimension (cf. \cite{point}). 
An efficient algorithm--considered as a nonlinear dynamical system--to generate {\it random} bistochastic matrices has
recently been presented \cite{CSBZ} (cf. \cite{griffiths,ZKSS}).

In the 1965-70 US 3,140-county migration study, a statistical 
test of Ling \cite{ling}  (designed 
for {\it undirected} graphs), based on the difference in
the ranks of two edges, was employed in a heuristic manner
\cite[pp. 7-8]{county}. 
For example, the 3,148th largest doubly-stochastic value, 0.12972 
(corresponding to the flow from Maui County to Hawaii County), {\it united}
the four counties of the state of Hawaii. The (considerably weaker) 
7,939th largest value, 
0.07340 (the link from Kauai County, Hawaii, to Nome, Alaska), {\it integrated}
the four-county 
state of Hawaii into a much larger 
2,464-county cluster. (Data for the additional [fifth] very small county of 
Kalawao were only given in the 1995-2000 analysis.) The difference of these
two ranks, 4,192 =  7,340 - 3,148, is a measure of isolation (``survival
time'') of this state as a cluster. Reference to Table 1 in \cite{county} 
showed the significance of the state of 
Hawaii as a functional 
internal migration unit at the 0.01 level \cite[p. 7]{county}. 
(In the computation of this table, the approximation was used that
the number of edges in the relevant 
digraphs was a negligible proportion of all 
possible $3,140 \times 3,139$ edges.)
\subsection{Random digraphs}
Also, the possibility of
employing the {\it asymptotic} theory of random digraphs 
\cite{palasti,karonski} for statistical testing purposes was raised 
in \cite{county}. In this regard, it was necessary to 
consider the 38,815-{\it th} largest entry of the doubly-stochastic matrix 
to complete the hierarchical 
clustering of the 3,140 counties. The probability is 0.973469 that a 
random digraph with 3,140 nodes and 38,414 links is strongly connected
\cite[p. 361]{karonski},
 where
$0.973469=e^{-2 e^{-4.30917}}$,
and $38,814= 3140 (\log{3140} +4.30917)$. 
Evidence of systematic structure in the migration
flows can, thus, be adduced, since the digraph based on the 38,814 greatest-valued links was {\it not} strongly connected \cite[p. 8]{county} 
(cf. \cite{killough}). (For our 1995-2000 analysis [Sec.~\ref{newer}], 
the counterpart of the probability 0.973469 is 0.107134.)

In a random digraph with a large number of nodes, the probability is close to 
one that all nodes are either isolated of lie in a single (``giant'') strong
component. The existence of intermediate-sized clusters is thus evidence of
non-randomness, even if such groups are not themselves 
significant according to the 
isolation (difference-of-ranks) criterion of Ling \cite{ling}. 
With randomly-generated data and many taxonomic units, one would expect
the two-stage procedure to yield a dendrogram exhibiting complete
chaining. So, although single-linkage clustering is often criticized
for producing chaining, chains can also be viewed 
simply as indications of inherent randomness in the data. 
In contrast to single-linkage clustering, strong component hierarchical
clustering can merge {\it more} than two clusters (children) into one
(parent) node. This serves to explain why fewer clusters (2,245) were
generated in the intercounty migration study than the 3,139 that 
single-linkage (in the absence of ties) would produce.
\subsection{A cluster-analytic isolation criterion}
Dubes and Jain \cite{DJ} provided ``a semi-tutorial review of the 
state-of-the-art 
in cluster validity, or the verification of results from clustering
algorithms''. Among other evaluative standards, they discussed isolation
criteria, which ``measure the distinctiveness or separation or gaps between a
cluster and its environment''.
Such a statistic was
developed and applied in \cite{qq2} in 
order to extract a small proportion of 5,385 clusters (3,140 of
them single units, 673 pairs, 230 triples, 104 quartets,\ldots) 
for detailed examination based on the 
two-stage analysis of the 1965-1970 United States intercounty migration 
table \cite{county}.

The largest value of the isolation criterion,
for all clusters of fewer than 2,940 units, was
attained by a region formed by the eight constituent counties of the state of
Connecticut. (Groups formed by the application of the
two-stage procedure to interareal migration data are, as a strong rule,
composed of contiguous areas \cite{qq,tree}.  This occurs even in the
absence of contiguity constraints, reflecting the distance decay of migration.)
The ll,080th largest doubly-standardized entry, 0.05666, corresponding to
movement from New Haven to (New York City suburban) 
Fairfield, unified these eight counties. Not until the 16,047{\it th}
largest doubly-standardized value, 0.04085 (the functional linkage 
from Litchfield, Connecticut  
to Berkshire, Massachusetts),
viewing the clustering procedure as an agglomerative one, was Connecticut
absorbed into a larger region.
The isolation criterion ($i$) for Connecticut is set equal to
\begin{equation}
i= 25.3175 
= - \log{\Big[\Big( (8 \times 7 + 3132 \times 3131)/(3140 \times 3139) \Big)^{(16047-11080)}\Big]}
\end{equation}
The term in large parentheses is the proportion of cells in the $3,140 
\times 3,140$ table associated
with either movement within (8-county) 
Connecticut or within the set of 3,132 
complementary
counties (since intracounty flows are not available, a diagonal
correction is made). This term, raised to the power shown, is the probability
(unadjusted for occupied cells) that {\it none} of $4,967 = 16,047 - 11,080$ 
consecutive
doubly-standardized values would correspond to movement between
Connecticut and its complement. (In our 1995-2000 analysis [Sec.~\ref{newer}],
we find analogously the result, 3,132= 12,107-8,975, 
yielding $i=16.1339$, still the most
signficant of any of the fifty-two 8-county clusters there.) 
Such a Connecticut-complement linkage
could possibly result in a merger: an {\it unobserved} phenomenon. 
(For further details, including maps, discussion  and extensive 
applications of the isolation criterion developed 
to the U. S. intercounty analysis,
see \cite{qq2}.) This isolation score ($i$) for the cluster formed by
the four counties of Hawaii--discussed above--was 12.21, 
while the District of Columbia had the highest score, 
23.81, for any single county \cite[Table I]{qq2}.
\section{Two-Stage Analyses of U. S. 
Intercounty Migration Flows} \label{newer}
\begin{figure}
\includegraphics{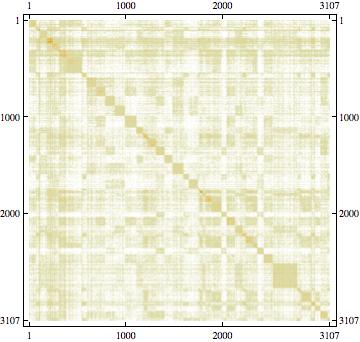}
\caption{\label{fig:matrixplotraw}Unadjusted 1995-2000 intercounty
U. S. migration table. The large square near the end--for 
alphabetical reasons--of the 
diagonal corresponds to the state with the most (254) counties, Texas.}
\end{figure}

\begin{figure}
\includegraphics{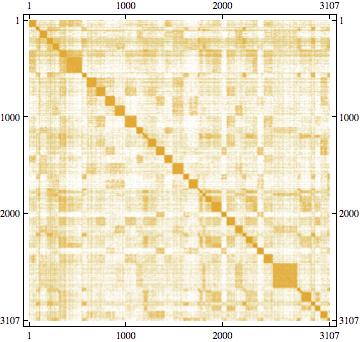}
\caption{\label{fig:matrixplotds}Doubly-stochastic form 
of the 1995-2000 intercounty
U. S. migration table}
\end{figure}

A $3,107 \times 3,107$ migration table for the United States 
for the period 1995-2000 can be
readily constructed from freely available data at the website,
http://www.census.gov/population/www/cen2000/ctytoctyflow/index.html. We 
have been able to conduct
a two-stage analysis of this table. 
In the 1965-70 analysis \cite{county}, 3,140 units of 3,141 had been 
utilized--with Loving County, TX, the smallest US county,  
being omitted since it had no recorded
in-migrants. The reduction to 3,107 units in the 1995-00 table is due
to the administrative 
amalgamation now of 34 independent cities of Virginia with 
neighboring counties. (Loving County is included in the later analysis, as
well as now the second smallest--and poorest--US county, Kalawao County, HI.)
The 1965-70 table was $94.5\%$ sparse (zero entries), and the 1995-00
table, $92.3\%$ sparse (the difference perhaps largely being due to 
the Census sampling design).
In Figs.~\ref{fig:matrixplotraw} and \ref{fig:matrixplotds}, we present
matrix plots of the unadjusted and adjusted tables. (The states are 
ordered alphabetically--not in terms of the postal [``zip''] code--and
the counties, alphabetically within states. 
County No. 1 is Autauga, AL; County No. 1000 
is Boyd, KY; County No. 2000 is Dunn, North Dakota; and County No. 
3,107 is Weston, WY.)
The largest square on the diagonal 
in both these figures corresponds to the state
with the most number of counties (254), that is, Texas.  
(The double-standardization--giving a more pronounced 
block-diagonal structure--brings out more strongly {\it intrastate} 
movements which would clearly tend to be favored over 
interstate ones due to effects 
of distance and possible state loyalties and ties.)
\subsection{Most cosmopolitan units} \label{most}
\begin{figure}
\includegraphics{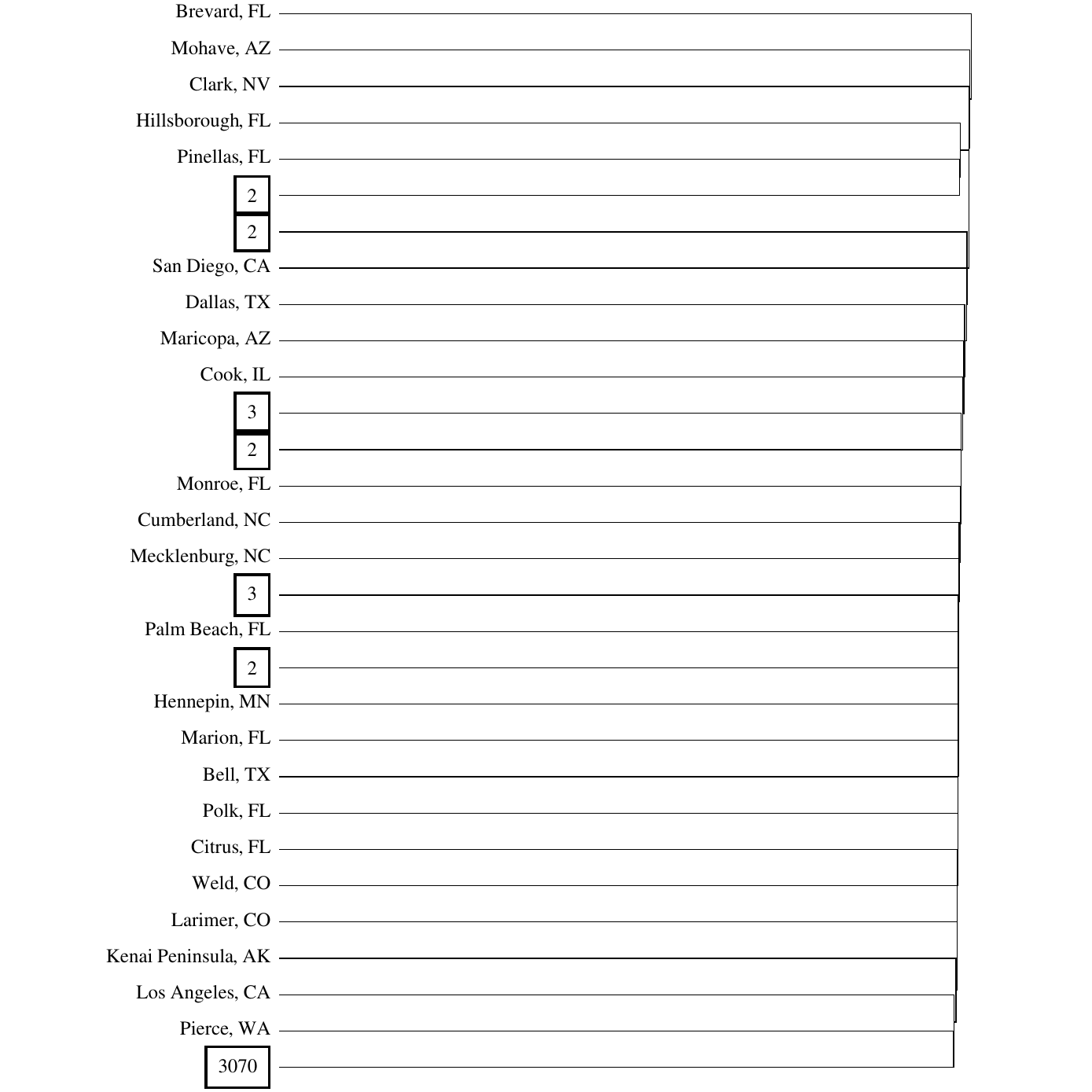}
\caption{\label{fig:dendrogram1}Truncated 
dendrogram--showing the most cosmopolitan
and groups of cosmopolitan counties--based upon doubly-standardized
1995-2000 intercounty migration flows. To obtain a distance-like
(dissimilarity) measure, we subtract the doubly-stochastic values 
from the largest such value, 0.530385}
\end{figure}

\begin{figure}[!ht]
\includegraphics{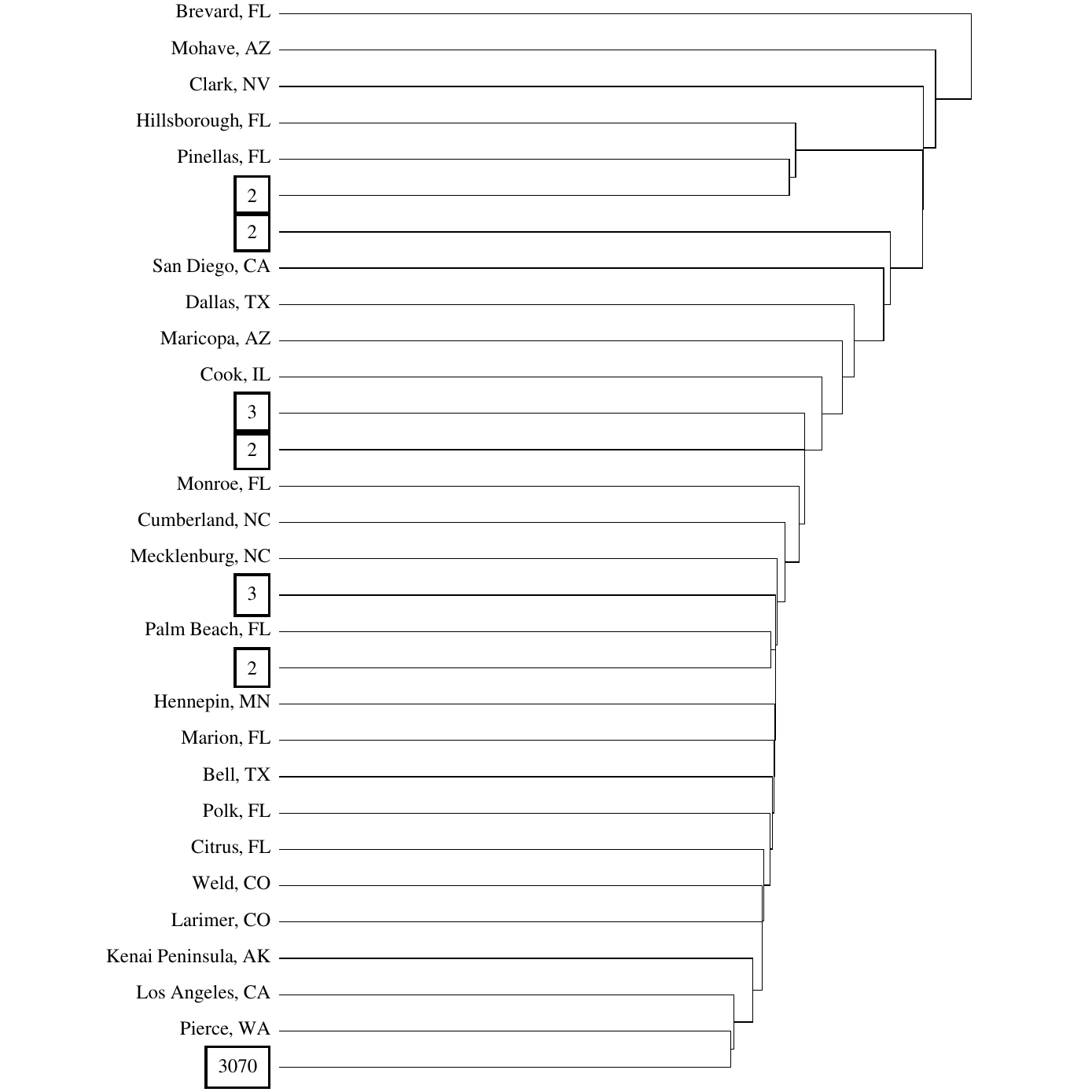}
\caption{\label{fig:dendrogram2}Truncated
dendrogram--showing the most cosmopolitan
and groups of cosmopolitan counties--based upon 
ordinal rankings of doubly-standardized
1995-2000 intercounty migration flows}
\end{figure}

In Fig.~\ref{fig:dendrogram1} we show the most cosmopolitan counties or
groups of counties based on the doubly-standardized values themselves,
while in 
(the less flat) 
Fig.~\ref{fig:dendrogram2}, 
the ordinal rank of the doubly-standardized value is used instead.
The doubly-standardized values associated with the evolution 
from beginning to end of
the hierarchical clustering range from 0.530385 to 0.0225427, while 
the ranks extend from 7 to 25,329. The comparable statistics for
the 1965-70 analysis based on 3,140 units 
were 0.47730 to 0.01659 and 24 to 38,815. So, ignoring any 
possibly necessary 
 corrections due to the slightly different sizes (3,140 {\it vs.} 
3,107) in the two periods 
and different degrees of sparsity ($94.5\%$ {\it vs.} 
$92.5\%$), one might 
conclude--since $0.0225427 > 0.01659$--that the most cosmopolitan
counties in the earlier analysis were more so (less "provincial") than the most cosmopolitan
counties in the later period.
(For the choice of most appropriate locations at which to truncate
dendrograms so as to distinguish cosmopolitan from provincial
units, see Sec.~\ref{provincial}.) 

\begin{figure}
\includegraphics{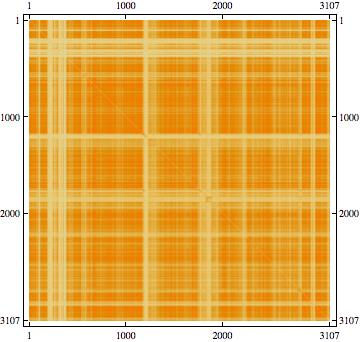}
\caption{\label{fig:ultrametric}Ultrametric fit to the 
1995-2000 doubly-stochastic internal migration flow table 
(Fig.~\ref{fig:matrixplotds})}
\end{figure}

\begin{figure}
\includegraphics{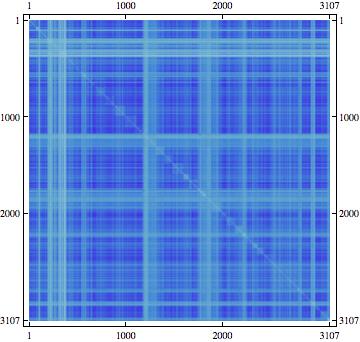}
\caption{\label{fig:ultrametricfit}
Residuals (mostly negative) 
from the fit of the ultrametric structure (Fig.~\ref{fig:ultrametric}) 
to the
1995-2000 doubly-stochastic table internal migration flow table 
(Fig.~\ref{fig:matrixplotds})}
\end{figure}

\begin{figure}
\includegraphics{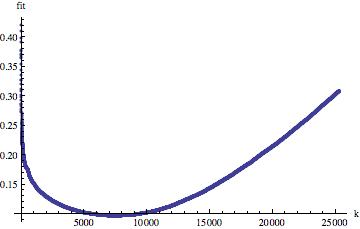}
\caption{\label{fig:clusteringfit}Goodness-of-fit of 
the ultrametric structure 
(Fig.~\ref{fig:ultrametric}) to the $k$ largest 1995-2000 
doubly-stochastic values. The best fit, 0.0964163, 
is attained at rank 7,229, corresponding to a doubly-stochastic
value of 0.0766761. At that threshold, there are 388 clusters 
(strong components).}
\end{figure}

The {\it non}-truncated 
(master) versions of these 
two figures (along with their counterparts based on 
the [smoothed] {\it second} power or square of the doubly-stochastic table) 
are given in the Electronic-only material  
and will be examined in Sec.~\ref{welldefined}.
In Fig.~\ref{fig:ultrametric} we show the {\it ultrametric fit} to
the doubly-stochastic table generated by the hierarchical clustering
procedure, and in Fig.~\ref{fig:ultrametricfit}, the {\it residuals} from
this fit. As a measure of goodness-of-fit, let us take the ratio of the 
sum of squares of the residuals 
from the largest $k=25,329$ entries (the number needed to 
complete the hierarchical 
clustering process) to the sum of squares of the 25,329 entries themselves.
This ratio was 0.30861. In Fig.~\ref{fig:clusteringfit}, we show
this measure of fit as a function of $k$. The minimum 
(best fit) of 0.0964163 is 
reached for the 7,229-{\it th} largest doubly-stochastic entry, 
0.0766761.

The list of most cosmopolitan
counties is much more ``Sunbelt''-oriented in nature than
in the 1965-70 analysis \cite{county} discussed above. 
(Let us note a technical point: even though our strong component
hierarchical clustering procedure can and often does unite more than
two smaller clusters, in order 
to fit within the Mathematica hierarchical 
clustering framework, we have to map our results into an
equivalent  hierarchical
clustering in which only binary mergers occur--though these may 
occur at equal thresholds. In our actual 1995-00 clustering, there were
2,497 mergers, as opposed to 3,106. The comparable figures for 1965-70 
were 2,245 and 3,139.)

The leading cosmopolitan counties found (and some of their 
apparently migration-relevant features), 
in decreasing order, are:

(1) Brevard, FL (the ``Space Coast'', the Kennedy Space Center);

(2) Mohave, AZ (Lake Havasu, Grand Canyon);

(3) Clark, NV (Las Vegas);

(4) Hillsborough and Pinellas, FL, which are grouped with 
     the pair  (represented by the topmost box with ``2'' 
     inside it), Pasco and 
     Hernando, FL. (This 
     quartet--having an isolation index of 11.9717--is completely 
     coterminous with the governmentally designated 
     Tampa-St. Petersburg-Clearwater Metropolitan 
     Statistical Area [MSA]. Additionally, 
     Pasco and Hernando have the greatest isolation
     index, 14.6413, of any pair in the entire analysis 
     [(Table~\ref{tab:table2}]);

(5) The next lower box with ``2'' in it, stands for the 
    southern Gulf Coast dyad formed by
    Collier County (East Naples) and Lee County (Fort Myers, a single-county 
    MSA),  FL;

(6) San Diego, CA;

(7) Dallas, TX;

(8) Maricopa, AZ (Phoenix);

(9) Cook, IL (Chicago);

(10) Orange, Seminole, and Osceola, FL (corresponding to 
     the upper box with ``3'' in it) 
     (these three counties, along with Lake County, form the Orlando-Kissimmee
      MSA);

(12) Sumter and Lake, FL (the next box with ``2'' in it);

(13) Monroe, FL (Key West);

(14) Cumberland, NC (giant Fort Bragg and Pope Air Force Base);

(15) Mecklenburg, NC (Charlotte);

(16) Martin, St. Lucie and Indian River, FL (the lower box 
     containing ``3'') (Indian River borders Brevard 
     County, the most cosmopolitan nationally);

(17) Palm Beach, FL together with the pair
      Miami-Dade and Broward, FL (the lowest box with ``2'' in it);
      (this southeastern Florida 
      triad comprises the Miami-Fort Lauderdale-Pompano Beach
      MSA, highlighted in gray in the master dendrograms [Electronic-only material]);

(18) Hennepin, MN (Minneapolis);

(19) Marion, FL (bordering the (17) cluster on the north);

(20) Bell, TX (Fort Hood);

(21) Polk, FL (Lakeland);

(22) Citrus, FL (formerly part of Hernando County);

(23) Weld, CO (Greeley);

(24) Larimer, CO (Fort Collins);

(25) Kenai Peninsula, AK (Seward);

(26) Los Angeles, CA; and

(27) Pierce, WA (Fort Lewis and McChord Air Force Base).

The comparable list for the earlier period 1965-70 takes the form
\cite[Table 1]{fields}: (1) Cook and DuPage, IL; (2) District of Columbia; (3) Dade and Broward, FL; (4) Pierce, WA; (5) Harris, TX (Houston); (6) Riverside and San Bernadino, CA; (7) Orange, CA; (8) Lake, IL (lying in the Chicago metropolitan area); (9) Monroe, FL; (10) Los Angeles, CA; (11) Pinellas, FL; (12) Brevard, FL; (13) Polk, FL; (14) Pulaski, Mo (Fort Leonard Wood); (15) Geary, KS (Fort Riley); (17) Wayne, MI (Detroit);
(18) Bell, TX; (19) Hillsborough, FL; (20) El Paso, CO (Air Force Academy); (21) Ventura, CA; (22) Cumberland, NC; (23) St. Louis County and City, MO;
(24) Norfolk, VA (Atlantic Fleet headquarters); (25) Arlington 
County and Alexandria City, VA; and 
(26) Sedgwick, KS (Wichita, McConnell Air
Force Base).

We see that this 1965-70 list is relatively weaker than the 1995-00 list
in terms of Sunbelt counties, but relatively stronger in terms of
counties with large military installations (though Brevard County, Florida
does have Patrick Air Force Base).
We note, in particular, 
that the first non-Sunbelt county in the 1965-70 list (that is, Cook, IL)
is ninth here, but (coupled with DuPage, IL) was most cosmopolitan
in the earlier analysis.  Also, the District of Columbia 
(colored pink in 
the associated 
master dendrograms [Electronic-only material], p. 3), which had the 
lowest threshold of isolation of any single county in the 1965-70 analysis
\cite[p. 31]{county}, {\it slips} very substantially.

The most immediate 
explanation for the relative decrease in cosmopolitanism of 
counties with large military installations would appear to be the
elimination of the draft in 1973--so, it would seem, 
military installations became
less relatively populated by transient, recently migrant individuals 
(draftees)--as well as the downsizing of the military since the Vietnam War. 
(The peak of 2.4 million troops was reached in 1969, while in 
2000, there were some 1.384 million military personnel.)
\subsection{Migration regions} \label{welldefined}
\subsubsection{Selected features}
We find--in the first two (searchable) master 
(non-truncated) dendrograms (Electronic-only material)--that 
the states of Hawaii (red, $i= 14.121$, p. 2), Connecticut (blue, 
$i=16.1339$, p. 2) 
and Rhode Island (green, $i=11.8384$, p. 3)
are reconstituted from their respective counties. (The most cosmopolitan
county in Hawaii is Honolulu, and in Connecticut, Fairfield [a ``bedroom
suburb''] of New York City. Both Hawaii and Connecticut
emerged as clusters in the 1965-70 analysis, while all the counties
of Rhode Island, but for historic Newport, were grouped.)
In both analyses, the fifteen southern counties
(colored black) of Maine are clustered (p. 8). 
(The northernmost, omitted county,
Aroostook  is agricultural, Canadian-oriented, and 
well-recognized as highly 
anomalous in terms of the general character 
of Maine \cite[p. 48, pp. 118-119]{county}.) 
In the 1995-00 dendrograms (Electronic-only material), these fifteen counties immediately merge
with six of the ten Lake Region counties of New Hampshire. The five 
counties of Rhode Island are also strongly linked with seven (or eight)
Massachusetts counties (Table I).
 
The five Pennsylvania counties (colored orange) of the Philadelphia-Camden-Wilmington  MSA 
(p. 5) were grouped in both analyses, and similarly the four 
New York metropolitan counties 
(colored brown) of Long Island (p. 2). 
(Their isolation indices in the 1995-2000
analyses are relatively weak, that is, 3.55119 and 2.50869, respectively.)

In the 1965-70 analysis, the dyad 
forming first in the agglomerative process was comprised
of the South Dakota counties of Dewey and Ziebach (p. 17), 
which together form the 
Cheyenne Indian Reservation. It is the sixth such couple
in the later analysis, with the four pairs now 
forming first being: (1) Stewart and Webster, Georgia (p. 17) 
(these two counties will be found to have the largest 
associated {\it diagonal} entries
when the doubly-stochastic table is {\it squared} in Sec.~\ref{square}; (2) 
Garfield and Petroleum, 
Montana (p. 32); (3) the Eastern Shore of Virginia, that is, 
Accomack and Northampton Counties (p. 3 and 
Table I); (4) and 
Cassia and Minidoka Counties, Idaho,
which form the Burley Micropolitan Statistical Area (p. 2 and 
Table I). Further, the 
{\it interstate} Jackson
Micropolitan Statistical Area--formed by Teton County, Idaho and Teton 
County, Wyoming--also comprises a strongly bound pair (p. 5). 
Our master dendrogams end with the pair of Alabama counties--lying 
in the Montgomery MSA--of  Autauga and Elmore.

The (strongly black-populated) 
Mississippi Delta is defined by Wikipedia as consisting 
of seventeen counties. 
Thirteen of these counties can be 
found in a certain fifteen-county 1995-00 cluster (colored magenta, 
having $i=2.8686$, p. 23). 
(In the 1965-70 analysis, we noted a six-county subcluster 
\cite[p. 57]{county}.)  The {\it southernmost} 
member of the seventeen-county
group, Warren County (Vicksburg), is omitted from the thirteen-member
cluster (along with Washington, Carroll and Holmes Counties).

The San Joaquin Valley of California is defined by Wikipedia as
comprised of seven counties. These seven, plus Madera County, are 
clustered (light green, p. 6). 

The six California counties that form the North Coast American
Viticultural Area also function as a migration region
(light brown, p. 3).

The three New York counties (Chautauqua, Cattaraugus and Allegany) 
forming the ``Southern Tier'' are highlighted in light blue (p. 7).
\subsubsection{Most well-defined 1995-2000 migration regions}
``South Jersey'' is--according to Wikipedia--composed of eight 
New Jersey counties. With the omission of its most northern 
member (in fact, classified for governmental 
purposes as in the New York metropolitan area), Ocean County, the
seven (Philadelphia-oriented) counties 
form a {\it very} well-defined migration region (colored light orange, 
$i=28.7301$, p. 5, while for 1965-70, $i=20.8996$ \cite[p. 64]{county} 
\cite{qq2}).
In fact, arranged in terms of decreasing values of $i$ for the period
1995-2000, this region emerges as the most well-defined in the entire
analysis (Tables I-II). (Many of the values of $i$ given in the tables
for the 1965-70 period
are available in \cite{qq2}.) Since all the values of $i$ for 1995-2000
listed are larger than $-\log{\frac{1}{100000}} = 11.5129$, we 
can infer that all these regions are significant at the 0.00001 level. 
\begin{table} \label{tab:table1}
\begin{tabular}{r| r | r |  c | c | c}
Region & States & Page & no. counties & i (1995-00) & i (1965-70) \\
\hline
\hline
South Jersey & NJ &6 & 7 &28.7301  & 20.8996  \\
Glades + Hendry + Okeechobee &FL & 1 & 3 & 23.474 &  \\
``Delmar'' + Baltimore & DE,MD & 5 & 15 & 20.283 & \\
Western Ohio + Randolph, IN &OH,IN & 25  & 14 & 20.0938 & \\
Western New York & NY & 7 & 18 & 19.4948 & \\
Rhode Island + S. E. Mass. & RI,MA & 3 & 12 &  18.6991 & \\
Greater Orlando& FL & 1 & 3 & 17.6523 & \\
Northern Lower Michigan & MI &8,9 & 26 & 17.2098 & \\
French Louisiana & LA &30,31 & 27 & 16.7764 & \\
Brevard & FL & 1 & 1 & 16.3097 & 19.6942 \\
Golden Triangle (Beaumont +) &TX& 4 & 6 & 16.1803 \\
Connecticut & CT &2 & 8 & 16.1339 & 25.3175 \\
Mohave (Kingman) & AZ & 1 & 1 & 15.463 & 6.39121 \\
Clark (Las Vegas) & NV &1  &1 & 15.1784 & 6.23128 \\
Rexburg, ID + Jackson, WY MSAs & ID,WY & 5 & 4 & 15.0882&  \\
Eastern Rust Belt &NJ,OH,PA,WV&24 &82 & 15.0412 & \\
Burley MSA & ID & 2  & 2 & 14.8809 & \\
Pasco + Hernando & FL &1 &2 & 14.6413 & \\
San Diego & CA & 1 & 1 & 14.2408  & 12.5938 \\
Maysville MSA + 3 counties & KY &19 & 5 & 14.1822 & \\
Hawaii & HI & 2 & 5\footnotemark[1]  & 14.121 & 12.21 \\
Northern High Plains & MT,ND,NE,SD&36,37&55&13.8799& \\
Middle Ohio Valley & IN,KY&24,25&27&13.821 \\
Eastern Shore & VA & 3 &2 & 13.7051 & \\
\hline
\hline
\end{tabular}
\caption{Most well-defined 1995-2000 migration regions and their isolation indices}
\footnotetext[1]{A fifth county, Kalawao, was included in the 
1995-00 data, but not in 1965-70}
\end{table}
\begin{table} \label{tab:table2}
\begin{tabular}{r| r | r |  c | c | c}
Region & States & Page & no. counties & i (1995-00) & i (1965-70) \\
\hline
\hline
Dallas & TX & 1 & 1 & 13.5473 & 14.8557 \\
Maine + 7 NH counties & ME,NH & 8  & 22 & 13.4716 & \\
Southeastern  Arizona & AZ& 2 &3 & 13.3503 & \\
Maricopa (Phoenix) & AZ & 1 & 1 & 13.2608 & 12.5479 \\
Eastern Upstate New York & NY & 7 & 28 & 13.3052 & \\
Michigan Thumb & MI & 6 & 6 & 13.2208 & \\
Wasatch Back &UT &11 &8& 13.1616 & \\
N. Vermont + Coos, NH &NH,VT& 11&10&13.0778& \\
S. Central Tennessee & TN & 22  & 10 & 13.3092 & \\
Northeast South Carolina& SC & 15  & 8 & 13.0276 & \\
Northern New England & MA,ME,NH,VT&9,10&42 & 12.8446 & \\
Cook (Chicago) & IL & 1 & 1 & 12.7682 & 16.8933 \\
Southeastern Indiana & IN & 25 & 10 & 12.7172 & \\
Northwestern Lower Michigan& MI & 9,10& 9 & 12.6567 & \\
High Colorado Rockies & CO & 3 & 3 & 12.5892 & \\
Joplin Area & MO & 5  & 3 & 12.3071 & \\
Central Savannah River & GA & 22 & 4 & 12.2086 & \\
Southern Maryland & MD & 3  & 3 & 12.1217 & \\
Amarillo (Potter + Randall) & TX & 1  &2 & 12.0528 & 8.16948 \\
Tampa MSA & FL & 1 &4 & 11.9717 & \\
York+Adams & PA & 3& 2 & 11.9433 & 13.7789 \\
Lake + Sumter & FL & 1  & 2 & 11.8635 & \\
Rhode Island & RI &  3  & 5 & 11.8384 & 11.7668\footnotemark[1]\\
Central Appalachia & MD,NC,TN,VA,WV&27,28&77& 11.7459 & \\
\hline
\hline
\end{tabular}
\caption{Most well-defined 1995-2000 migration regions and their isolation indices (cont.)}
\footnotetext[1]{Newport County was not directly clustered with the other four 
counties of the state in 1965-70}
\end{table}

{\it French Louisiana} is defined by Wikipedia as the amalgamation of
Acadiana/''Cajun Country'' 
(22 parishes) and Greater New Orleans (7 parishes)--St. Charles 
and St. John the Baptist being common to both--giving a 27-parish region.
Our analysis yields a well-defined ($i=16.7764$) 27-parish region also,
having 24 parishes in common with the Wikipedia definition.
Our candidate region contains the three parishes, Allen, 
(the Mississippi-bordering pair of) East Feliciana and 
West Feliciana, but lacks those of Avoyelles (immediately adjacent, 
however, in the dendrogram), Orleans 
(coextensive with the City of New Orleans) and St. Tammany (also in
the New Orleans metropolitan area). (The last two parishes--located on pp. 
9 and 15 of the first two master 
dendrograms--are relatively cosmopolitan--as might be anticipated from their
wide [pre-Hurricane Katrina] renown.)

The Northern New England region is composed of the three states of Maine,
New Hampshire and Vermont, plus the two (mutually well-separated)
Massachusetts counties of (western) Berkshire and (northeastern) Essex.

Our Northern Lower Michigan region is composed of twenty-six counties,
twenty-two of which are contained in the twenty-seven county Wikipedia
definition. Our region, however,  extends further to the Southeast around 
Saginaw Bay, with Isabella,
Midland, Bay and Saginaw counties, and omits five of Wikepedia's
southwesternly situated  ones
(Wexford, Missaukee, Osceola, Lake and Mason).

Adams County, PA was created from part of York County, PA (Table II).

We did omit from Table I the rather anomalous 
twelve-county, four-state (ID, OR, UT, WA) cluster found
on p. 19 of the dendrograms, even though it has $i = 18.1505$. 
It is essentially composed of two 
rather remote {\it noncontiguous} (OR-WA and ID-UT) sets of areas
united by 
the links Clark, ID $\rightarrow$ Sherman, OR (having a 
doubly-stochastic value of 0.270946, the 261-{\it st} largest) 
and Skamania, WA $\rightarrow$ 
Bear Lake, ID (0.135147, the 2,655-{\it th} largest). (We have no
immediate explanation for these apparently surprisingly relatively
large values.)
\subsubsection{Cosmopolitan/Provincial Boundary} \label{provincial}
There is a large 2,423-county cluster ({\it lacking} all of the 
New England and Hawaiian counties)--having the high value 
$i=27.7726$--stretching in the dendrogram from 
Navarro, TX (p. 9) until the very end (Autauga, AL). 
(This might be considered to be a  domain 
of {\it lesser} cosmpolitan counties or groups of counties.) 
It is a subcluster of 
a 2,588-county cluster ($i=27.7304$)
stretching from Quay, NM (p. 7), again to the end.
Still larger, but somewhat weaker,  is a 3,069-county cluster, $i= 20.2582$, 
extending from Salt Lake, UT [ p. 1] until the end.
Further, starting with Wayne, NC, but excluding Androscoggin, ME 
(p. 8), there is a 2,483-county  
cluster extending to the end with $i = 19.7518$. 
If we were to maps such results, the cosmopolitan counties--it would
seem--would comprise ``archipelagos'' in the ``sea'' of 
provincial counties.
\subsection{Master dendrograms based on the {\bf square} 
of the doubly-stochastic table} \label{square}
\begin{figure}
\includegraphics{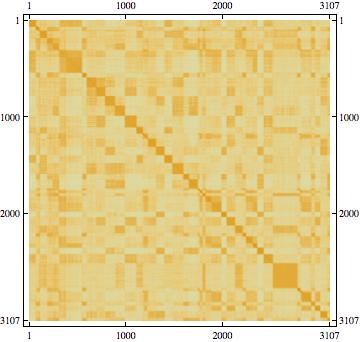}
\caption{\label{fig:matrixplotSquare}Square of the doubly-stochastic form
(Fig.~\ref{fig:matrixplotds})
of the 1995-2000 intercounty U. S. migration table. Only $2.82\%$
of the entries of the matrix are 0, while $92.3\%$ are in the 
unsquared matrix}
\end{figure}

\begin{figure}
\includegraphics{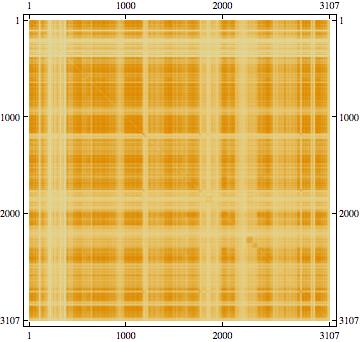}
\caption{\label{fig:ultrametricSquare}Ultrametric fit to the {\it square} 
of the
1995-2000 doubly-stochastic internal 
migration flow table (Fig.~\ref{fig:matrixplotSquare})}
\end{figure}

\begin{figure}
\includegraphics{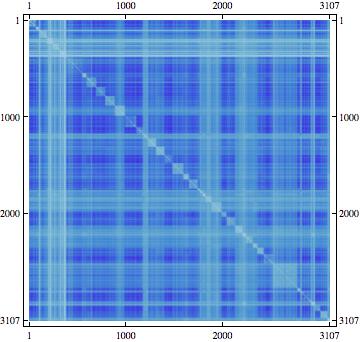}
\caption{\label{fig:ultrametricfitsquare}
Residuals (mostly negative)
of the fit of the ultrametric structure 
(Fig.~\ref{fig:ultrametricSquare}) to the {\it square} of the 
1995-2000 doubly-stochastic table internal migration flow table 
(Fig.~\ref{fig:matrixplotSquare})}
\end{figure}

\begin{figure}
\includegraphics{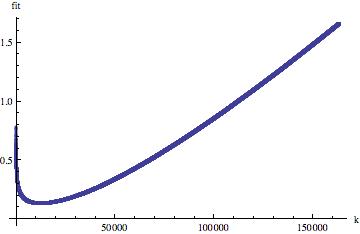}
\caption{\label{fig:clusteringfitSquare}Goodness-of-fit of
the ultrametric structure to the $k$ largest values in the {\it square} of 
the  1995-2000
doubly-stochastic table. The best fit, 0.130718,
is attained at rank 12,603, corresponding to a doubly-stochastic
value of 0.0206195. At that threshold, there are 377 clusters 
(strong components).}
\end{figure}

Matrix multiplying 
the doubly-stochastic form of the 1995-2000 (zero-diagonal) 
intercounty migration table by itself, we obtain another doubly-stochastic 
table (Fig.~\ref{fig:matrixplotSquare}),
but now one with non-zero diagonal entries--which ranged from a high
of 0.314164 and 0.3060604 for the members of a pair of 
small Georgia counties--Webster and Stewart (we recall that
these two counties were the first cluster formed in 
the hierarchical [agglomerative] process--to lows of 0 and 0.0000253087 
for the Hawaii counties of Kalawao and Kauai, respectively. 
(Kalawao County--once the site of a leprosy colony--was 
in many respects anomalous because of its very 
small size, and might in retrospect been readily omitted from the 
analyses. 
Of course, if a county has a large associated diagonal entry in 
the newly derived doubly-stochastic table, 
its off-diagonal entries, which are the only ones which 
affect its clustering properties, 
will tend to be reduced in size.) 
The resulting master dendrograms (presented in the Electronic-only 
material, along with the ultrametric [ordinal] fit in 
Fig.~\ref{fig:ultrametricSquare}, the residuals from this fit 
in Fig.~\ref{fig:ultrametricfitsquare}
and the goodness-of-fit measure in 
Fig.~\ref{fig:clusteringfitSquare})--again 
employing the strong component hierarchical 
clustering methodology--are 
even more biased in cosmopolitanism to Sunbelt counties. 
(The largest 163,341 doubly-stochastic values  were required to 
complete the strong component hierarchical clustering, much more than
the 25,329 needed in the original [unsquared] analysis.)

There are obviously many interesting 
significant features in these figures, as the 
many long strings of counties within single states, 
apparent upon examination,  would indicate. 
The isolation index for the seven-county ``South Jersey'' migration region
(p. 7) has now climbed to 49.8337, while the five clustered counties of Hawaii 
(p. 12) have an index of 34.4687. There is a still 
higher  index of 64.2316 for a 39-county
region (pp. 6-7) composed of all the counties 
of Maine, New Hampshire and Vermont, but for 
Bennington 
County, VT (p. 6), which is adjacent to New York and 
clustered with counties of that (non-New England) 
state instead.  This 39-county tri-state
region is included along with Eastern Upstate 
New York counties in a 66-county region
(with a very high 
$i=118.237$). There was now also a 7-county Connecticut region
(lacking New York City ``bedroom''-suburban 
Fairfield County), having $i=32.913$ 
(p. 5).

Additionally, to identify some of the other prominent clusters, 
we have a 15-county New York region 
($i=34.0272$, p. 13), a 12-county Arkansas region ($i=32.9012$, pp. 8-9),
a 20-county (northeastern) 
North Carolina region ($i=32.5555$, pp. 14-15), 
a 32-county Ohio region ($i=29.8344$, pp. 18-19),
a 16-county (western) 
South Carolina region ($i=29.9287$, p. 
13), an 83-county NJ-NY-PA-WV (``Rust Belt'') 
region ($i=26.9231$, pp. 12-13), a 17-county tri-state ``Delmarva'' 
region ($i=26.1837$, p. 13), and 
a joining of seven northern 
Florida counties with seventy-seven of Georgia ($i= 24.5441$, pp. 29-30).
\subsection{Use of teleporting random walk}
Motivated by the widespread interest in and emulation of the PageRank
algorithm used by popular search engines such as Google 
\cite{brinpage,langville},
we took a weighted combination of the doubly-stochastic 1995-2000
U. S. intercounty migration table and the zero-diagonal $3,107 \times 3,107$ 
doubly-stochastic table with all its off-diagonal entries equal to
$\frac{1}{3106}$. A weight of 0.9 was applied to the former table, and 0.1
to the latter table.

Again, precisely (the largest) 
25,329 entries of the resultant table were needed to complete the (strong
component) hierarchical clustering process.
We noted that ``South Jersey'' was again a 7-county cluster 
with precisely the same isolation index of 28.7301 (Table I).
Overall, our original clustering appeared to be totally 
robust in its qualitative features 
to the effect of the teleporting random walk, at least
with the particular weights (0.9,0.1) we employed. Potentially, 
if the {\it square} of the doubly-stochastic table were similarly teleported, 
the associated hierarchical clustering results might not be so robust.
\section{Concluding Remarks}
\subsection{Aggregation issues}
One might--using the indicated two-stage 
procedure--compare the hierarchical structure of geographic 
areas using internal migration tables at {\it different} levels of 
geographic aggregation
(counties, states, regions...) (cf. \cite{point}).
To again use the example of France, based on a 
$21 \times 21$ interregional
table for 1962-68, R{\'e}gion Parisienne was the most hub-like 
\cite[Sec. 4.1]{tree} \cite{france}, while using a finer
$89 \times 89$ interdepartmental table for 1954-62, the dyad 
composed of Seine 
(that is Paris and its immediate suburbs) together with the 
encircling
Seine-et-Oise (administratively eliminated in 1964) 
was most cosmopolitan \cite{winchester} 
\cite[Sec. 6.1]{tree}. 
(In \cite{point}, `` two distinct approaches to assessing 
the effect of geographic scale on spatial interactions'' were 
developed.)

We, in fact, can directly compare the results of our U. S. 1965-70 
migration between 3,140 counties study \cite{county} with a 
highly detailed study 
\cite{SEAS} for the very same period
conducted on the more aggregate level of 510 State Economic Areas 
(SEAs, collections of counties). In terms of relative cosmopolitan
characteristics, the list based on the SEAs does have some different
emphases than that given above in terms of the counties. According
to Fig. 2 of \cite{SEAS}, the most cosmopolitan SEAs, in decreasing order,
were Alaska; Hawaii (2 SEAs); Southeast Florida (3 
SEAs); Southwest Florida; North Florida;
the Chicago SMSA; the New York SMSA; Norfolk-Portsmouth SMSA; San Bernadino
and Riverside SMSA; the District of Columbia; and the Maryland suburbs of 
D. C. (2 SEAs). 
(As previously noted, in the county-level 1965-70 analysis, the 
two most cosmopolitan entities were the Chicago metropolitan pair of
Cook and DuPage, and the District of Columbia--while in the 1995-00
intercounty analysis, Brevard, FL and Mohave, AZ played these roles.)
Let us also bring to the reader's attention a 
2005 discussion paper in which the 1995-2000
U. S. inter{\it state} migration table is studied using 
both double-standardization and ``social network analysis''
\cite{maier}. (Hierarchical clustering is also employed, but apparently
not that form based on strong components.)
\subsection{Max-flow/Min-cut application}
In \cite{newman1}, Newman applied the famous Ford-Fulkerson 
{\it max-flow/min-cut theorem} \cite[Chap. 22]{nijenhuis} to
weighted networks (which he mapped onto unweighted {\it multigraphs}). Earlier, this theorem had been used to study 
Spanish \cite{multiterminal}, Philippine \cite{philippine}, and 
Brazilian, Mexican and Argentinian \cite{brazil} 
internal migration, 
US interindustry flows \cite[pp. 18-28]{tree2} \cite{IO2} \cite[Sec. III]{gosling} and the international flow of college students \cite{seps} 
(cf. \cite{cor})--all the corresponding flows now being
left unadjusted, that is {\it not} (doubly- nor singly-) standardized.

In this ``multiterminal'' 
approach, the maximum flow and the dual minimum edge cut-sets, 
between {\it all} ordered pairs of nodes are found. Those cuts 
(often few or even {\it null} in number) which partition the 
$N$ nodes nontrivially--that is, into two sets each of cardinality greater than
1--are noted. The set in each such pair with the fewer nodes is regarded
as a nodal cluster (region, in the geographic context). It has the 
interesting, defining property that fewer people migrate into (from) it, as 
a whole, than into (from) its node. In the Spanish context, the 
(nodal) province of 
Badajoz was found to have a particularly large out-migration sphere of
influence, and 
the (Basque) province of 
Vizcaya (site of Bilbao and Guernica), 
an extensive in-migration field \cite{multiterminal}.
In an analysis of 1967 US interindustry transactions based on 468 
industries, among the industries functioning as  nodes
of  {\it production} complexes with large 
numbers of members were: Advertising; 
Blast Furnaces and Steel Mills; Electronic
Components; and Paperboard Containers and Boxes. Conversely, among those 
serving as nodes of {\it consumption} complexes 
were Petroleum Refining and Meat Animals 
\cite{IO2,gosling}.
\subsection{Subdominant eigenvalue}
\begin{figure}
\includegraphics{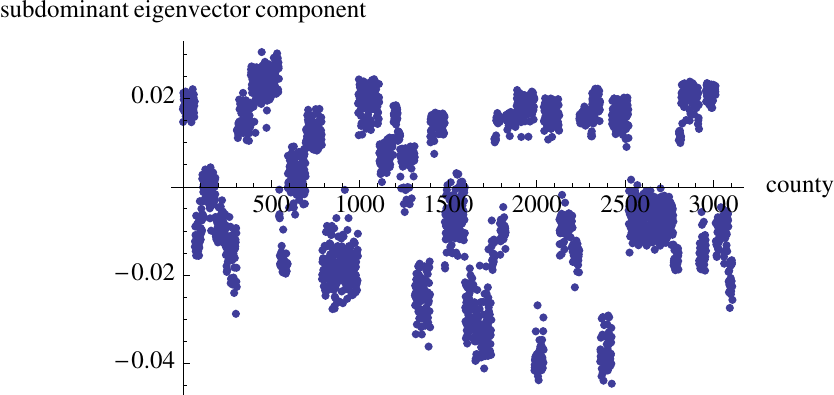}
\caption{\label{fig:subdominant}Components of the subdominant eigenvector
of the doubly-stochastic form of the 1995-2000 U. S. intercounty
migration table. The highest-situated evident cluster is composed
of counties of the state of Georgia}
\end{figure}

Pentney and Meila have extended spectral clustering algorithms to 
``asymmetric affinities'' \cite{pentney}. In line with their approach,
we computed the subdominant eigenvalue (0.906253) of 
the $3,107 \times 3,107$ doubly-stochastic 1995-2000 intercounty
migration table, and the associated eigenvector. 
(Trivially, the {\it dominant} eigenvalue is 1, and the components of
the corresponding eigenvector all equal.)  
Interestingly,
the largest (most positive) seventy-eight components of this vector {\it all}
corresponded to counties of Georgia, while the smallest (most negative)
one hundred and ten components were all from one or another of the
contiguous triad of Great Plains states, North Dakota, South Dakota and
Nebraska. 
(The most negative two
values are for Dewey and Ziebach Counties of South Dakota, which as we
have previously indicated form the Cheyenne Indian Reservation, and 
was the first cluster to form in the 1965-70 [agglomerative] 
hierarchical clustering \cite{county}.) 
In Fig.~\ref{fig:subdominant} we present a ``list plot''
of these components. (The counties are listed alphabetically within states, 
and the states themselves alphabetically also.) 
A rough gestalt estimate might yield 
some thirty to
forty clusters.
Dorogovtsev and Mendes have reviewed ``the recent rapid progress in 
the statistical physics of evolving networks'' \cite{evolution}.
\subsection{U. S. internal migration network}
We have presented (Electronic material) 
master dendrograms descriptive of the rich geographical and
sociological evolving tapestry of the United States--as reflected in
the 1965-1970 and 1995-2000 migration flows between the 3,000+ 
county-level units. Our results have been 
derived using a demonstratedly-insightful
two-stage methodology--double-standardization of the recorded flows 
followed by
(strong component) hierarchical clustering--applicable
to (weighted, directed) socioeconomic networks, in general.
Applying a graph-theoretic isolation 
criterion, we extracted particularly
distinct large multicounty migration regions, 
well describable as ``French Louisiana'', ``Northern Lower 
Michigan'', ``Northern New England'', {\it et al}.
Certain tightly-knit
functional clusters--for example, the states of Connecticut, Hawaii, 
as well as
``South Jersey''--are invariant over the thirty-year study period.
Broad ``cosmopolitan'' or ``hub-like'' migration to and from 
``Sunbelt'' counties (Clark County, Nevada [Las Vegas], for instance)
became relatively more conspicuous and migration associated with
counties with large military installations (Pierce County, Washington 
[Fort Lewis and McChord Air Force Base], for example), 
less so. Further, 
the most cosmopolitan units for 1965-70
(the paired Chicago metropolitan 
counties of Cook and DuPage, Illinois, and the 
District of Columbia heading the list) were more cosmopolitan 
in character than the leading ones in the later analysis. 
We supplemented  these analyses by studying both the {\it square} 
and a ``teleported random walk'' form 
of the doubly-stochastic table, as well as its subdominant 
eigenvalue. The hierarchical clustering obtained is robust
against teleportation.

\begin{acknowledgments}
I would like to express appreciation to the Kavli Institute for Theoretical Physics (KITP) for technical support, as well as Waldo Tobler for granting 
permission  to use Figures 1-4.
\end{acknowledgments}

\bibliography{Evolving10}

\begin{thebibliography}{109}
\expandafter\ifx\csname natexlab\endcsname\relax\def\natexlab#1{#1}\fi
\expandafter\ifx\csname bibnamefont\endcsname\relax
  \def\bibnamefont#1{#1}\fi
\expandafter\ifx\csname bibfnamefont\endcsname\relax
  \def\bibfnamefont#1{#1}\fi
\expandafter\ifx\csname citenamefont\endcsname\relax
  \def\citenamefont#1{#1}\fi
\expandafter\ifx\csname url\endcsname\relax
  \def\url#1{\texttt{#1}}\fi
\expandafter\ifx\csname urlprefix\endcsname\relax\def\urlprefix{URL }\fi
\providecommand{\bibinfo}[2]{#2}
\providecommand{\eprint}[2][]{\url{#2}}

\bibitem[{\citenamefont{Barab{\'a}si}(2003)}]{linked}
\bibinfo{author}{\bibfnamefont{A.-L.} \bibnamefont{Barab{\'a}si}},
  \emph{\bibinfo{title}{Linked: How everything is connected to everything else
  and what it means for business, science, and everyday life}}
  (\bibinfo{publisher}{Plume}, \bibinfo{address}{New York},
  \bibinfo{year}{2003}).

\bibitem[{\citenamefont{Siegfried}(2006)}]{siegfried}
\bibinfo{author}{\bibfnamefont{T.}~\bibnamefont{Siegfried}},
  \emph{\bibinfo{title}{A beautiful math: John Nash, game theory, and the
  modern quest for a code of nature}} (\bibinfo{publisher}{Joseph Henry},
  \bibinfo{address}{Washington}, \bibinfo{year}{2006}).

\bibitem[{\citenamefont{Slater}(1984{\natexlab{a}})}]{tree}
\bibinfo{author}{\bibfnamefont{P.~B.} \bibnamefont{Slater}},
  \emph{\bibinfo{title}{Tree representations of internal migration flows and
  related topics}} (\bibinfo{publisher}{Community and Organization Res. Inst.},
  \bibinfo{address}{Santa Barbara}, \bibinfo{year}{1984}{\natexlab{a}}).

\bibitem[{\citenamefont{Dubes}(1985)}]{dubes}
\bibinfo{author}{\bibfnamefont{R.~C.} \bibnamefont{Dubes}},
  \bibinfo{journal}{J. Classif.} \textbf{\bibinfo{volume}{2}},
  \bibinfo{pages}{141} (\bibinfo{year}{1985}).

\bibitem[{\citenamefont{Slater}(1976{\natexlab{a}})}]{japan}
\bibinfo{author}{\bibfnamefont{P.~B.} \bibnamefont{Slater}},
  \bibinfo{journal}{Regional Stud.} \textbf{\bibinfo{volume}{10}},
  \bibinfo{pages}{123} (\bibinfo{year}{1976}{\natexlab{a}}).

\bibitem[{\citenamefont{Slater}(1976{\natexlab{b}})}]{france}
\bibinfo{author}{\bibfnamefont{P.~B.} \bibnamefont{Slater}},
  \bibinfo{journal}{IEEE Syst. Man. Cyb.} \textbf{\bibinfo{volume}{6}},
  \bibinfo{pages}{321} (\bibinfo{year}{1976}{\natexlab{b}}).

\bibitem[{\citenamefont{Slater and Winchester}(1978)}]{winchester}
\bibinfo{author}{\bibfnamefont{P.~B.} \bibnamefont{Slater}} \bibnamefont{and}
  \bibinfo{author}{\bibfnamefont{H.~L.~M.} \bibnamefont{Winchester}},
  \bibinfo{journal}{IEEE Syst. Man. Cyb.} \textbf{\bibinfo{volume}{8}},
  \bibinfo{pages}{635} (\bibinfo{year}{1978}).

\bibitem[{\citenamefont{Slater}(1983{\natexlab{a}})}]{science}
\bibinfo{author}{\bibfnamefont{P.~B.} \bibnamefont{Slater}},
  \bibinfo{journal}{Scientometrics} \textbf{\bibinfo{volume}{5}},
  \bibinfo{pages}{55} (\bibinfo{year}{1983}{\natexlab{a}}).

\bibitem[{\citenamefont{Slater}(1984{\natexlab{b}})}]{partial}
\bibinfo{author}{\bibfnamefont{P.~B.} \bibnamefont{Slater}},
  \bibinfo{journal}{Environ. Plann. A} \textbf{\bibinfo{volume}{16}},
  \bibinfo{pages}{545} (\bibinfo{year}{1984}{\natexlab{b}}).

\bibitem[{\citenamefont{Slater}(1977{\natexlab{a}})}]{IO}
\bibinfo{author}{\bibfnamefont{P.~B.} \bibnamefont{Slater}},
  \bibinfo{journal}{Empirical Econ.} \textbf{\bibinfo{volume}{2}},
  \bibinfo{pages}{1} (\bibinfo{year}{1977}{\natexlab{a}}).

\bibitem[{\citenamefont{Slater}(1976{\natexlab{c}})}]{college}
\bibinfo{author}{\bibfnamefont{P.~B.} \bibnamefont{Slater}},
  \bibinfo{journal}{Res. Higher Educ.} \textbf{\bibinfo{volume}{4}},
  \bibinfo{pages}{305} (\bibinfo{year}{1976}{\natexlab{c}}).

\bibitem[{\citenamefont{Gentileschi and Slater}(1980)}]{gentileschi}
\bibinfo{author}{\bibfnamefont{M.~L.} \bibnamefont{Gentileschi}}
  \bibnamefont{and} \bibinfo{author}{\bibfnamefont{P.~B.}
  \bibnamefont{Slater}}, \bibinfo{journal}{Riv. Geog. Ital.}
  \textbf{\bibinfo{volume}{87}}, \bibinfo{pages}{133} (\bibinfo{year}{1980}).

\bibitem[{\citenamefont{Slater}(1975{\natexlab{a}})}]{metron}
\bibinfo{author}{\bibfnamefont{P.~B.} \bibnamefont{Slater}},
  \bibinfo{journal}{Metron} \textbf{\bibinfo{volume}{33}}, \bibinfo{pages}{182}
  (\bibinfo{year}{1975}{\natexlab{a}}).

\bibitem[{\citenamefont{Slater}(1976{\natexlab{d}})}]{SEAS}
\bibinfo{author}{\bibfnamefont{P.~B.} \bibnamefont{Slater}},
  \bibinfo{journal}{Rev. Public Data Use} \textbf{\bibinfo{volume}{4}},
  \bibinfo{pages}{32} (\bibinfo{year}{1976}{\natexlab{d}}).

\bibitem[{\citenamefont{Slater}(1981{\natexlab{a}})}]{qq}
\bibinfo{author}{\bibfnamefont{P.~B.} \bibnamefont{Slater}},
  \bibinfo{journal}{Quality and Quantity} \textbf{\bibinfo{volume}{15}},
  \bibinfo{pages}{179} (\bibinfo{year}{1981}{\natexlab{a}}).

\bibitem[{\citenamefont{Slater}(1976{\natexlab{e}})}]{spain}
\bibinfo{author}{\bibfnamefont{P.~B.} \bibnamefont{Slater}},
  \bibinfo{journal}{Trabajos de Estadistica y de Investigaci{\'o}n Operativa}
  \textbf{\bibinfo{volume}{27}}, \bibinfo{pages}{175}
  (\bibinfo{year}{1976}{\natexlab{e}}).

\bibitem[{\citenamefont{Slater and Schwarz}(1979)}]{schwarz}
\bibinfo{author}{\bibfnamefont{P.~B.} \bibnamefont{Slater}} \bibnamefont{and}
  \bibinfo{author}{\bibfnamefont{W.}~\bibnamefont{Schwarz}},
  \bibinfo{journal}{IEEE Syst. Man Cyber.} \textbf{\bibinfo{volume}{9}},
  \bibinfo{pages}{381} (\bibinfo{year}{1979}).

\bibitem[{\citenamefont{Slater}(1979{\natexlab{a}})}]{romania}
\bibinfo{author}{\bibfnamefont{P.~B.} \bibnamefont{Slater}},
  \bibinfo{journal}{Econ. Computation and Econ. Cyber}
  \textbf{\bibinfo{volume}{13}}, \bibinfo{pages}{97}
  (\bibinfo{year}{1979}{\natexlab{a}}).

\bibitem[{\citenamefont{Slater}(1975{\natexlab{b}})}]{russia}
\bibinfo{author}{\bibfnamefont{P.~B.} \bibnamefont{Slater}},
  \bibinfo{journal}{Soviet Geog.} \textbf{\bibinfo{volume}{16}},
  \bibinfo{pages}{453} (\bibinfo{year}{1975}{\natexlab{b}}).

\bibitem[{\citenamefont{Hirst and Slater}(1976)}]{hirst1}
\bibinfo{author}{\bibfnamefont{M.~A.} \bibnamefont{Hirst}} \bibnamefont{and}
  \bibinfo{author}{\bibfnamefont{P.~B.} \bibnamefont{Slater}},
  \bibinfo{journal}{E. Afr. Geog. Rev.} \textbf{\bibinfo{volume}{13}},
  \bibinfo{pages}{9} (\bibinfo{year}{1976}).

\bibitem[{\citenamefont{Slater}(1981{\natexlab{b}})}]{seps}
\bibinfo{author}{\bibfnamefont{P.~B.} \bibnamefont{Slater}},
  \bibinfo{journal}{Socio-Econ. Plann. Sci.} \textbf{\bibinfo{volume}{15}},
  \bibinfo{pages}{1} (\bibinfo{year}{1981}{\natexlab{b}}).

\bibitem[{\citenamefont{Slater}(1976{\natexlab{f}})}]{india}
\bibinfo{author}{\bibfnamefont{P.~B.} \bibnamefont{Slater}},
  \bibinfo{journal}{The Geographer} \textbf{\bibinfo{volume}{23}},
  \bibinfo{pages}{1} (\bibinfo{year}{1976}{\natexlab{f}}).

\bibitem[{\citenamefont{Slater}(1983{\natexlab{b}})}]{county}
\bibinfo{author}{\bibfnamefont{P.~B.} \bibnamefont{Slater}},
  \emph{\bibinfo{title}{Migration regions of the United States: two
  county-level 1965-70 analyses}} (\bibinfo{publisher}{Community and
  Organization Res. Inst.}, \bibinfo{address}{Santa Barbara},
  \bibinfo{year}{1983}{\natexlab{b}}).

\bibitem[{\citenamefont{Slater}(1986)}]{tree2}
\bibinfo{author}{\bibfnamefont{P.~B.} \bibnamefont{Slater}},
  \emph{\bibinfo{title}{Large scale data analytic studies in the social
  sciences}} (\bibinfo{publisher}{Community and Organization Res. Inst.},
  \bibinfo{address}{Santa Barbara}, \bibinfo{year}{1986}).

\bibitem[{\citenamefont{Masser and Scheurwater}(1980)}]{masser}
\bibinfo{author}{\bibfnamefont{I.}~\bibnamefont{Masser}} \bibnamefont{and}
  \bibinfo{author}{\bibfnamefont{J.~S.} \bibnamefont{Scheurwater}},
  \bibinfo{journal}{Environ. Plann. A} \textbf{\bibinfo{volume}{12}},
  \bibinfo{pages}{1357} (\bibinfo{year}{1980}).

\bibitem[{\citenamefont{Kim}(1984)}]{kim}
\bibinfo{author}{\bibfnamefont{K.~H.} \bibnamefont{Kim}},
  \bibinfo{journal}{Math. Soc. Sci.} \textbf{\bibinfo{volume}{8}},
  \bibinfo{pages}{195} (\bibinfo{year}{1984}).

\bibitem[{\citenamefont{Hirst}(1976)}]{hirst2}
\bibinfo{author}{\bibfnamefont{M.~A.} \bibnamefont{Hirst}},
  \bibinfo{journal}{Environ. Plann. A} \textbf{\bibinfo{volume}{9}},
  \bibinfo{pages}{99} (\bibinfo{year}{1976}).

\bibitem[{\citenamefont{Findlay and Slater}(1981)}]{findlay}
\bibinfo{author}{\bibfnamefont{A.}~\bibnamefont{Findlay}} \bibnamefont{and}
  \bibinfo{author}{\bibfnamefont{P.~B.} \bibnamefont{Slater}},
  \bibinfo{journal}{Environ. Plann. A} \textbf{\bibinfo{volume}{13}},
  \bibinfo{pages}{645} (\bibinfo{year}{1981}).

\bibitem[{\citenamefont{Holmes and Slater}(1977)}]{holmes1}
\bibinfo{author}{\bibfnamefont{J.~H.} \bibnamefont{Holmes}} \bibnamefont{and}
  \bibinfo{author}{\bibfnamefont{P.~B.} \bibnamefont{Slater}},
  \bibinfo{journal}{IEEE Trans. Syst. Man Cyber.} \textbf{\bibinfo{volume}{7}},
  \bibinfo{pages}{474} (\bibinfo{year}{1977}).

\bibitem[{\citenamefont{Holmes and Slater}(1978)}]{holmes2}
\bibinfo{author}{\bibfnamefont{J.~H.} \bibnamefont{Holmes}} \bibnamefont{and}
  \bibinfo{author}{\bibfnamefont{P.~B.} \bibnamefont{Slater}},
  \bibinfo{journal}{IEEE Trans. Syst. Man Cyber.} \textbf{\bibinfo{volume}{8}},
  \bibinfo{pages}{325} (\bibinfo{year}{1978}).

\bibitem[{\citenamefont{Holmes}(1978)}]{holmes3}
\bibinfo{author}{\bibfnamefont{J.~H.} \bibnamefont{Holmes}},
  \bibinfo{journal}{Prog. Human Geog.} \textbf{\bibinfo{volume}{2}},
  \bibinfo{pages}{467} (\bibinfo{year}{1978}).

\bibitem[{\citenamefont{Fischer et~al.}(1993)\citenamefont{Fischer,
  Essletzbichler, Gassler, and Tricht}}]{telephone}
\bibinfo{author}{\bibfnamefont{M.}~\bibnamefont{Fischer}},
  \bibinfo{author}{\bibfnamefont{J.}~\bibnamefont{Essletzbichler}},
  \bibinfo{author}{\bibfnamefont{J.}~\bibnamefont{Gassler}}, \bibnamefont{and}
  \bibinfo{author}{\bibfnamefont{G.}~\bibnamefont{Tricht}},
  \bibinfo{journal}{Geog. Anal.} \textbf{\bibinfo{volume}{25}},
  \bibinfo{pages}{224} (\bibinfo{year}{1993}).

\bibitem[{\citenamefont{Baumann et~al.}(1983)\citenamefont{Baumann, Fischer,
  and Schubert}}]{baumann}
\bibinfo{author}{\bibfnamefont{J.~H.} \bibnamefont{Baumann}},
  \bibinfo{author}{\bibfnamefont{M.~M.} \bibnamefont{Fischer}},
  \bibnamefont{and} \bibinfo{author}{\bibfnamefont{U.}~\bibnamefont{Schubert}},
  \bibinfo{journal}{Papers in Regional Sci.} \textbf{\bibinfo{volume}{52}},
  \bibinfo{pages}{53} (\bibinfo{year}{1983}).

\bibitem[{\citenamefont{Clark}(1982)}]{clark}
\bibinfo{author}{\bibfnamefont{G.~L.} \bibnamefont{Clark}},
  \bibinfo{journal}{Environ. Plann. A} \textbf{\bibinfo{volume}{14}},
  \bibinfo{pages}{145} (\bibinfo{year}{1982}).

\bibitem[{\citenamefont{Boyd}(1980)}]{boyd}
\bibinfo{author}{\bibfnamefont{J.}~\bibnamefont{Boyd}}, \bibinfo{journal}{IEEE
  Syst. Man Cyber.} \textbf{\bibinfo{volume}{10}}, \bibinfo{pages}{101}
  (\bibinfo{year}{1980}).

\bibitem[{\citenamefont{Pandit}(1994)}]{pandit}
\bibinfo{author}{\bibfnamefont{K.}~\bibnamefont{Pandit}},
  \bibinfo{journal}{Profess. Geog.} \textbf{\bibinfo{volume}{46}},
  \bibinfo{pages}{331} (\bibinfo{year}{1994}).

\bibitem[{\citenamefont{Hoover and Giarrantani}(1984)}]{hoover}
\bibinfo{author}{\bibfnamefont{E.~M.} \bibnamefont{Hoover}} \bibnamefont{and}
  \bibinfo{author}{\bibfnamefont{F.}~\bibnamefont{Giarrantani}},
  \emph{\bibinfo{title}{An introduction to regional economics}}
  (\bibinfo{publisher}{Knopf}, \bibinfo{address}{New York},
  \bibinfo{year}{1984}).

\bibitem[{\citenamefont{Slater}()}]{discernment}
\bibinfo{author}{\bibfnamefont{P.~B.} \bibnamefont{Slater}},
  \emph{\bibinfo{title}{Discernment of hubs and clusters in socioeconomic
  networks}}, \eprint{arXiv:0807.1550}.

\bibitem[{\citenamefont{Takayuki and Yoshida}(205)}]{takayuki}
\bibinfo{author}{\bibfnamefont{S.}~\bibnamefont{Takayuki}} \bibnamefont{and}
  \bibinfo{author}{\bibfnamefont{H.}~\bibnamefont{Yoshida}},
  \emph{\bibinfo{title}{Data analysis of asymmetric structures: advanced
  approaches in computational statistics}} (\bibinfo{publisher}{Marcel Dekker},
  \bibinfo{address}{New York}, \bibinfo{year}{205}).

\bibitem[{\citenamefont{Noronha and Goodchild}(1992)}]{noronha}
\bibinfo{author}{\bibfnamefont{V.~T.} \bibnamefont{Noronha}} \bibnamefont{and}
  \bibinfo{author}{\bibfnamefont{M.~F.} \bibnamefont{Goodchild}},
  \bibinfo{journal}{Annals Assoc. Amer. Geog.} \textbf{\bibinfo{volume}{82}},
  \bibinfo{pages}{86} (\bibinfo{year}{1992}).

\bibitem[{\citenamefont{C{\"o}rvers et~al.}(2008)\citenamefont{C{\"o}rvers,
  Hensen, and Bongaerts}}]{corvers}
\bibinfo{author}{\bibfnamefont{F.}~\bibnamefont{C{\"o}rvers}},
  \bibinfo{author}{\bibfnamefont{M.}~\bibnamefont{Hensen}}, \bibnamefont{and}
  \bibinfo{author}{\bibfnamefont{D.}~\bibnamefont{Bongaerts}},
  \bibinfo{journal}{Reg. Stud.} \textbf{\bibinfo{volume}{42}},
  \bibinfo{pages}{DOI:10.1080/00343400701654103} (\bibinfo{year}{2008}).

\bibitem[{\citenamefont{Fienberg}(1970)}]{fienberg}
\bibinfo{author}{\bibfnamefont{S.}~\bibnamefont{Fienberg}},
  \bibinfo{journal}{Ann. Math. Stat.} \textbf{\bibinfo{volume}{41}},
  \bibinfo{pages}{907} (\bibinfo{year}{1970}).

\bibitem[{\citenamefont{Bacharach}(1970)}]{bacharach}
\bibinfo{author}{\bibfnamefont{M.~A.} \bibnamefont{Bacharach}},
  \emph{\bibinfo{title}{Biproportional matrices and input-output change}}
  (\bibinfo{publisher}{Cambridge Univ.}, \bibinfo{address}{Cambridge},
  \bibinfo{year}{1970}).

\bibitem[{\citenamefont{Mosteller}(1968)}]{mosteller}
\bibinfo{author}{\bibfnamefont{F.}~\bibnamefont{Mosteller}},
  \bibinfo{journal}{J. Amer. Statist. Assoc.} \textbf{\bibinfo{volume}{63}},
  \bibinfo{pages}{1} (\bibinfo{year}{1968}).

\bibitem[{\citenamefont{Louck}(1997)}]{louck}
\bibinfo{author}{\bibfnamefont{J.~D.} \bibnamefont{Louck}},
  \bibinfo{journal}{Found. Phys.} \textbf{\bibinfo{volume}{27}},
  \bibinfo{pages}{1085} (\bibinfo{year}{1997}).

\bibitem[{\citenamefont{Cappellini et~al.}()\citenamefont{Cappellini, Sommers,
  Bruzda, and \.Zyczkowski}}]{CSBZ}
\bibinfo{author}{\bibfnamefont{V.}~\bibnamefont{Cappellini}},
  \bibinfo{author}{\bibfnamefont{H.-J.} \bibnamefont{Sommers}},
  \bibinfo{author}{\bibfnamefont{W.}~\bibnamefont{Bruzda}}, \bibnamefont{and}
  \bibinfo{author}{\bibfnamefont{K.}~\bibnamefont{\.Zyczkowski}},
  \emph{\bibinfo{title}{Nonlinear dynamics in constructing random bistochastic
  matrices}}, \eprint{arXiv:0711.3345}.

\bibitem[{\citenamefont{Bengtsson}()}]{unistochastic}
\bibinfo{author}{\bibfnamefont{I.}~\bibnamefont{Bengtsson}},
  \emph{\bibinfo{title}{The importance of being unistochastic}},
  \eprint{quant-ph/0403088}.

\bibitem[{\citenamefont{Romney}(1971)}]{romney}
\bibinfo{author}{\bibfnamefont{A.~K.} \bibnamefont{Romney}}, in
  \emph{\bibinfo{booktitle}{Measuring endogamy}} (\bibinfo{publisher}{MIT
  Press}, \bibinfo{address}{Cambridge}, \bibinfo{year}{1971}), pp.
  \bibinfo{pages}{191--213}.

\bibitem[{\citenamefont{Wong}(1992)}]{wong}
\bibinfo{author}{\bibfnamefont{D.~W.~S.} \bibnamefont{Wong}},
  \bibinfo{journal}{Profess. Geog.} \textbf{\bibinfo{volume}{44}},
  \bibinfo{pages}{340} (\bibinfo{year}{1992}).

\bibitem[{\citenamefont{Eriksson}(1980)}]{eriksson}
\bibinfo{author}{\bibfnamefont{J.}~\bibnamefont{Eriksson}},
  \bibinfo{journal}{Math. Program.} \textbf{\bibinfo{volume}{18}},
  \bibinfo{pages}{146} (\bibinfo{year}{1980}).

\bibitem[{\citenamefont{Macgill}(1977)}]{macgill}
\bibinfo{author}{\bibfnamefont{S.~M.} \bibnamefont{Macgill}},
  \bibinfo{journal}{Environ. Plann. A} \textbf{\bibinfo{volume}{9}},
  \bibinfo{pages}{687} (\bibinfo{year}{1977}).

\bibitem[{\citenamefont{Parlett and Landis}(1982)}]{parlett}
\bibinfo{author}{\bibfnamefont{B.~N.} \bibnamefont{Parlett}} \bibnamefont{and}
  \bibinfo{author}{\bibfnamefont{T.~L.} \bibnamefont{Landis}},
  \bibinfo{journal}{Lin. Alg. Applics.} \textbf{\bibinfo{volume}{48}},
  \bibinfo{pages}{53} (\bibinfo{year}{1982}).

\bibitem[{\citenamefont{Sinkhorn and Knopp}(1967)}]{sinkhorn}
\bibinfo{author}{\bibfnamefont{R.}~\bibnamefont{Sinkhorn}} \bibnamefont{and}
  \bibinfo{author}{\bibfnamefont{P.}~\bibnamefont{Knopp}},
  \bibinfo{journal}{Pac. J. Math.} \textbf{\bibinfo{volume}{21}},
  \bibinfo{pages}{343} (\bibinfo{year}{1967}).

\bibitem[{\citenamefont{Mirsky}(1971)}]{mirsky}
\bibinfo{author}{\bibfnamefont{L.}~\bibnamefont{Mirsky}},
  \emph{\bibinfo{title}{Transversal Theory}} (\bibinfo{publisher}{Academic},
  \bibinfo{address}{New York}, \bibinfo{year}{1971}).

\bibitem[{\citenamefont{Linial et~al.}(2000)\citenamefont{Linial,
  Samorodnitsky, and Wigderson}}]{linial}
\bibinfo{author}{\bibfnamefont{N.}~\bibnamefont{Linial}},
  \bibinfo{author}{\bibfnamefont{A.}~\bibnamefont{Samorodnitsky}},
  \bibnamefont{and}
  \bibinfo{author}{\bibfnamefont{A.}~\bibnamefont{Wigderson}},
  \bibinfo{journal}{Combinatorica} \textbf{\bibinfo{volume}{20}},
  \bibinfo{pages}{545} (\bibinfo{year}{2000}).

\bibitem[{\citenamefont{Simonoff}(1995)}]{simonoff}
\bibinfo{author}{\bibfnamefont{J.~S.} \bibnamefont{Simonoff}},
  \bibinfo{journal}{J. Statist. Plann. Infer.} \textbf{\bibinfo{volume}{47}},
  \bibinfo{pages}{41} (\bibinfo{year}{1995}).

\bibitem[{\citenamefont{Slater}(1980{\natexlab{a}})}]{boundary}
\bibinfo{author}{\bibfnamefont{P.~B.} \bibnamefont{Slater}},
  \bibinfo{journal}{IEEE Syst. Man Cyber.} \textbf{\bibinfo{volume}{10}},
  \bibinfo{pages}{678} (\bibinfo{year}{1980}{\natexlab{a}}).

\bibitem[{\citenamefont{Brin and Page}(1998)}]{brinpage}
\bibinfo{author}{\bibfnamefont{S.}~\bibnamefont{Brin}} \bibnamefont{and}
  \bibinfo{author}{\bibfnamefont{L.}~\bibnamefont{Page}},
  \bibinfo{journal}{Comput. Netw. ISDN Systs.} \textbf{\bibinfo{volume}{30}},
  \bibinfo{pages}{107} (\bibinfo{year}{1998}).

\bibitem[{\citenamefont{Langville and Meyer}(2006)}]{langville}
\bibinfo{author}{\bibfnamefont{A.~N.} \bibnamefont{Langville}}
  \bibnamefont{and} \bibinfo{author}{\bibfnamefont{C.~D.} \bibnamefont{Meyer}},
  \emph{\bibinfo{title}{Google's PageRank and beyond: the science of search
  engines}} (\bibinfo{publisher}{Princeton Univ. Press},
  \bibinfo{address}{Princeton}, \bibinfo{year}{2006}).

\bibitem[{\citenamefont{Maier and Vyborny}(2005)}]{maier}
\bibinfo{author}{\bibfnamefont{G.}~\bibnamefont{Maier}} \bibnamefont{and}
  \bibinfo{author}{\bibfnamefont{M.}~\bibnamefont{Vyborny}},
  \emph{\bibinfo{title}{Internal migration between US states--a social network
  analysis}} (\bibinfo{publisher}{Dept. Urban Regional Develop.},
  \bibinfo{address}{WU-Wien}, \bibinfo{year}{2005}).

\bibitem[{\citenamefont{Gower and Ross}(1989)}]{gower1}
\bibinfo{author}{\bibfnamefont{J.~C.} \bibnamefont{Gower}} \bibnamefont{and}
  \bibinfo{author}{\bibfnamefont{G.~J.~S.} \bibnamefont{Ross}},
  \bibinfo{journal}{Appl. Stat.} \textbf{\bibinfo{volume}{18}},
  \bibinfo{pages}{54} (\bibinfo{year}{1989}).

\bibitem[{\citenamefont{Ozawa}(1983)}]{ozawa}
\bibinfo{author}{\bibfnamefont{K.}~\bibnamefont{Ozawa}},
  \bibinfo{journal}{Patt. Recog.} \textbf{\bibinfo{volume}{16}},
  \bibinfo{pages}{201} (\bibinfo{year}{1983}).

\bibitem[{\citenamefont{Hubert}(1973)}]{hubert}
\bibinfo{author}{\bibfnamefont{L.~J.} \bibnamefont{Hubert}},
  \bibinfo{journal}{Psychometrika} \textbf{\bibinfo{volume}{38}},
  \bibinfo{pages}{63} (\bibinfo{year}{1973}).

\bibitem[{\citenamefont{Leusmann}(1977)}]{leusmann}
\bibinfo{author}{\bibfnamefont{C.}~\bibnamefont{Leusmann}},
  \bibinfo{journal}{Comput. Applics.} \textbf{\bibinfo{volume}{769}},
  \bibinfo{pages}{769} (\bibinfo{year}{1977}).

\bibitem[{\citenamefont{Chilko}(1980)}]{chilko}
\bibinfo{author}{\bibfnamefont{D.}~\bibnamefont{Chilko}}, \bibinfo{journal}{SAS
  Supplemental Library User's Guide} pp. \bibinfo{pages}{65--70}
  (\bibinfo{year}{1980}).

\bibitem[{\citenamefont{Schwartz}(1982)}]{schwartz}
\bibinfo{author}{\bibfnamefont{J.}~\bibnamefont{Schwartz}},
  \bibinfo{journal}{Not. Amer. Math. Soc.} \textbf{\bibinfo{volume}{29}},
  \bibinfo{pages}{502} (\bibinfo{year}{1982}).

\bibitem[{\citenamefont{Tarjan}(1982)}]{tarjan}
\bibinfo{author}{\bibfnamefont{R.~E.} \bibnamefont{Tarjan}},
  \bibinfo{journal}{Info. Proc. Lett.} \textbf{\bibinfo{volume}{14}},
  \bibinfo{pages}{26} (\bibinfo{year}{1982}).

\bibitem[{\citenamefont{Tarjan}(1983)}]{tarjan2}
\bibinfo{author}{\bibfnamefont{R.~E.} \bibnamefont{Tarjan}},
  \bibinfo{journal}{Info. Proc. Lett.} \textbf{\bibinfo{volume}{17}},
  \bibinfo{pages}{37} (\bibinfo{year}{1983}).

\bibitem[{\citenamefont{Slater}(1987)}]{tarjanslater}
\bibinfo{author}{\bibfnamefont{P.~B.} \bibnamefont{Slater}},
  \bibinfo{journal}{Environ. Plann. A} \textbf{\bibinfo{volume}{19}},
  \bibinfo{pages}{117} (\bibinfo{year}{1987}).

\bibitem[{\citenamefont{Hartfiel and Spellman}(1972)}]{hartfiel}
\bibinfo{author}{\bibfnamefont{D.~F.} \bibnamefont{Hartfiel}} \bibnamefont{and}
  \bibinfo{author}{\bibfnamefont{J.~W.} \bibnamefont{Spellman}},
  \bibinfo{journal}{Proc. Amer. Math. Soc.} \textbf{\bibinfo{volume}{36}},
  \bibinfo{pages}{389} (\bibinfo{year}{1972}).

\bibitem[{\citenamefont{Hansen and Jaumard}(1997)}]{hansen}
\bibinfo{author}{\bibfnamefont{P.}~\bibnamefont{Hansen}} \bibnamefont{and}
  \bibinfo{author}{\bibfnamefont{B.}~\bibnamefont{Jaumard}},
  \bibinfo{journal}{Math. Programming} \textbf{\bibinfo{volume}{79}},
  \bibinfo{pages}{191} (\bibinfo{year}{1997}).

\bibitem[{\citenamefont{Clauset et~al.}()\citenamefont{Clauset, Moore, and
  Newman}}]{cmn}
\bibinfo{author}{\bibfnamefont{A.}~\bibnamefont{Clauset}},
  \bibinfo{author}{\bibfnamefont{C.}~\bibnamefont{Moore}}, \bibnamefont{and}
  \bibinfo{author}{\bibfnamefont{M.}~\bibnamefont{Newman}},
  \emph{\bibinfo{title}{Structural inference of hierarchies in networks}},
  \eprint{arXiv:physics/0610051}.

\bibitem[{\citenamefont{Mihaescu and Pachter}(2008)}]{radu}
\bibinfo{author}{\bibfnamefont{R.}~\bibnamefont{Mihaescu}} \bibnamefont{and}
  \bibinfo{author}{\bibfnamefont{L.}~\bibnamefont{Pachter}},
  \bibinfo{journal}{Proc. Natl. Acad. Sci.} \textbf{\bibinfo{volume}{105}},
  \bibinfo{pages}{13206} (\bibinfo{year}{2008}).

\bibitem[{\citenamefont{Johnson}(1967)}]{johnson}
\bibinfo{author}{\bibfnamefont{S.~C.} \bibnamefont{Johnson}},
  \bibinfo{journal}{Psychometrika} \textbf{\bibinfo{volume}{32}},
  \bibinfo{pages}{241} (\bibinfo{year}{1967}).

\bibitem[{\citenamefont{Rammal et~al.}(1986)\citenamefont{Rammal, Toulouse, and
  Virasoro}}]{rammal}
\bibinfo{author}{\bibfnamefont{R.}~\bibnamefont{Rammal}},
  \bibinfo{author}{\bibfnamefont{G.}~\bibnamefont{Toulouse}}, \bibnamefont{and}
  \bibinfo{author}{\bibfnamefont{M.~A.} \bibnamefont{Virasoro}},
  \bibinfo{journal}{Rev. Mod. Phys.} \textbf{\bibinfo{volume}{58}},
  \bibinfo{pages}{765} (\bibinfo{year}{1986}).

\bibitem[{\citenamefont{Costa}(2004)}]{costa}
\bibinfo{author}{\bibfnamefont{L.~F.} \bibnamefont{Costa}},
  \bibinfo{journal}{Phys. Rev. Lett.} \textbf{\bibinfo{volume}{93}},
  \bibinfo{pages}{098702} (\bibinfo{year}{2004}).

\bibitem[{\citenamefont{Costa}(2007)}]{costa2}
\bibinfo{author}{\bibfnamefont{L.~F.} \bibnamefont{Costa}},
  \bibinfo{journal}{New J. Phys.} \textbf{\bibinfo{volume}{9}},
  \bibinfo{pages}{311} (\bibinfo{year}{2007}).

\bibitem[{\citenamefont{Slater}(1978)}]{siegen}
\bibinfo{author}{\bibfnamefont{P.~B.} \bibnamefont{Slater}}, in
  \emph{\bibinfo{booktitle}{Competition among small regions}}
  (\bibinfo{publisher}{Nomos Verlagsgesellschaft},
  \bibinfo{address}{Baden-Baden}, \bibinfo{year}{1978}), pp.
  \bibinfo{pages}{63--70}.

\bibitem[{\citenamefont{Slater}(1982)}]{manila}
\bibinfo{author}{\bibfnamefont{P.~B.} \bibnamefont{Slater}},
  \bibinfo{journal}{GeoJournal} \textbf{\bibinfo{volume}{6}},
  \bibinfo{pages}{477} (\bibinfo{year}{1982}).

\bibitem[{\citenamefont{Slater}(1975{\natexlab{c}})}]{turkish}
\bibinfo{author}{\bibfnamefont{P.~B.} \bibnamefont{Slater}}, in
  \emph{\bibinfo{booktitle}{1975 Proceedings of the American Statistical
  Association}} (\bibinfo{year}{1975}{\natexlab{c}}), pp.
  \bibinfo{pages}{207--213}.

\bibitem[{\citenamefont{Slater}(1984{\natexlab{c}})}]{fields}
\bibinfo{author}{\bibfnamefont{P.~B.} \bibnamefont{Slater}},
  \bibinfo{journal}{Geog. Anal.} \textbf{\bibinfo{volume}{16}},
  \bibinfo{pages}{65} (\bibinfo{year}{1984}{\natexlab{c}}).

\bibitem[{\citenamefont{Slater}(1980{\natexlab{b}})}]{gosling}
\bibinfo{author}{\bibfnamefont{P.~B.} \bibnamefont{Slater}}, in
  \emph{\bibinfo{booktitle}{Input Output and Marketing}}
  (\bibinfo{publisher}{Augustus M. Kelley}, \bibinfo{address}{London},
  \bibinfo{year}{1980}{\natexlab{b}}), pp. \bibinfo{pages}{257--276}.

\bibitem[{\citenamefont{Rosvall and Bergstrom}(2008)}]{rosvall}
\bibinfo{author}{\bibfnamefont{M.}~\bibnamefont{Rosvall}} \bibnamefont{and}
  \bibinfo{author}{\bibfnamefont{C.~T.} \bibnamefont{Bergstrom}},
  \bibinfo{journal}{Proc. Natl. Acad. Sci.} \textbf{\bibinfo{volume}{105}},
  \bibinfo{pages}{1118} (\bibinfo{year}{2008}).

\bibitem[{\citenamefont{E et~al.}(2008)\citenamefont{E, Li, and
  Vanden-Eijnden}}]{e}
\bibinfo{author}{\bibfnamefont{W.}~\bibnamefont{E}},
  \bibinfo{author}{\bibfnamefont{T.}~\bibnamefont{Li}}, \bibnamefont{and}
  \bibinfo{author}{\bibfnamefont{E.}~\bibnamefont{Vanden-Eijnden}},
  \bibinfo{journal}{Proc. Natl. Acad. Sci} \textbf{\bibinfo{volume}{105}},
  \bibinfo{pages}{7907} (\bibinfo{year}{2008}).

\bibitem[{\citenamefont{Slater}(1976{\natexlab{g}})}]{multiterminal}
\bibinfo{author}{\bibfnamefont{P.~B.} \bibnamefont{Slater}},
  \bibinfo{journal}{Environ. Plann. A} \textbf{\bibinfo{volume}{8}},
  \bibinfo{pages}{875} (\bibinfo{year}{1976}{\natexlab{g}}).

\bibitem[{\citenamefont{Lin and Xie}(1998)}]{loglinear}
\bibinfo{author}{\bibfnamefont{G.}~\bibnamefont{Lin}} \bibnamefont{and}
  \bibinfo{author}{\bibfnamefont{Y.}~\bibnamefont{Xie}},
  \bibinfo{journal}{Amer. Sociol. Rev.} \textbf{\bibinfo{volume}{63}},
  \bibinfo{pages}{900} (\bibinfo{year}{1998}).

\bibitem[{\citenamefont{Price and Zubrzycki}(1962)}]{price}
\bibinfo{author}{\bibfnamefont{C.~A.} \bibnamefont{Price}} \bibnamefont{and}
  \bibinfo{author}{\bibfnamefont{J.}~\bibnamefont{Zubrzycki}},
  \bibinfo{journal}{Pop. Stud.} \textbf{\bibinfo{volume}{16}},
  \bibinfo{pages}{123} (\bibinfo{year}{1962}).

\bibitem[{\citenamefont{Bock}(1996)}]{bock}
\bibinfo{author}{\bibfnamefont{H.~H.} \bibnamefont{Bock}},
  \bibinfo{journal}{Comput. Stat. Data Anal.} \textbf{\bibinfo{volume}{23}},
  \bibinfo{pages}{5} (\bibinfo{year}{1996}).

\bibitem[{\citenamefont{Duncan}(1979)}]{duncan}
\bibinfo{author}{\bibfnamefont{O.~D.} \bibnamefont{Duncan}},
  \bibinfo{journal}{Amer. J. Sociol.} \textbf{\bibinfo{volume}{84}},
  \bibinfo{pages}{793} (\bibinfo{year}{1979}).

\bibitem[{\citenamefont{Rao and Sabavala}(1981)}]{rao}
\bibinfo{author}{\bibfnamefont{V.~R.} \bibnamefont{Rao}} \bibnamefont{and}
  \bibinfo{author}{\bibfnamefont{D.~J.} \bibnamefont{Sabavala}},
  \bibinfo{journal}{J. Consumer Res.} \textbf{\bibinfo{volume}{8}},
  \bibinfo{pages}{85} (\bibinfo{year}{1981}).

\bibitem[{\citenamefont{Blumstein and Larson}(1969)}]{blumstein}
\bibinfo{author}{\bibfnamefont{A.}~\bibnamefont{Blumstein}} \bibnamefont{and}
  \bibinfo{author}{\bibfnamefont{R.}~\bibnamefont{Larson}},
  \bibinfo{journal}{Operat. Res.} \textbf{\bibinfo{volume}{17}},
  \bibinfo{pages}{199} (\bibinfo{year}{1969}).

\bibitem[{\citenamefont{Rothkopf}(1957)}]{rothkopf}
\bibinfo{author}{\bibfnamefont{E.~Z.} \bibnamefont{Rothkopf}},
  \bibinfo{journal}{J. Experiment. Psych.} \textbf{\bibinfo{volume}{53}},
  \bibinfo{pages}{94} (\bibinfo{year}{1957}).

\bibitem[{\citenamefont{Slater}(1985{\natexlab{a}})}]{point}
\bibinfo{author}{\bibfnamefont{P.~B.} \bibnamefont{Slater}},
  \bibinfo{journal}{Environ. Plann. A} \textbf{\bibinfo{volume}{17}},
  \bibinfo{pages}{1025} (\bibinfo{year}{1985}{\natexlab{a}}).

\bibitem[{\citenamefont{Griffiths}(1974)}]{griffiths}
\bibinfo{author}{\bibfnamefont{R.~C.} \bibnamefont{Griffiths}},
  \bibinfo{journal}{Canad. J. Math.} \textbf{\bibinfo{volume}{26}},
  \bibinfo{pages}{600} (\bibinfo{year}{1974}).

\bibitem[{\citenamefont{{\.Z}yczkowski
  et~al.}(2003)\citenamefont{{\.Z}yczkowski, Ku{\'s}, S{\l}omczy{\'n}ski, and
  Sommers}}]{ZKSS}
\bibinfo{author}{\bibfnamefont{K.}~\bibnamefont{{\.Z}yczkowski}},
  \bibinfo{author}{\bibfnamefont{M.}~\bibnamefont{Ku{\'s}}},
  \bibinfo{author}{\bibfnamefont{W.}~\bibnamefont{S{\l}omczy{\'n}ski}},
  \bibnamefont{and} \bibinfo{author}{\bibfnamefont{H.-J.}
  \bibnamefont{Sommers}}, \bibinfo{journal}{J. Phys. A}
  \textbf{\bibinfo{volume}{36}}, \bibinfo{pages}{3425} (\bibinfo{year}{2003}).

\bibitem[{\citenamefont{Ling}(1973)}]{ling}
\bibinfo{author}{\bibfnamefont{R.~F.} \bibnamefont{Ling}}, \bibinfo{journal}{J.
  Amer. Statist. Assoc.} \textbf{\bibinfo{volume}{68}}, \bibinfo{pages}{159}
  (\bibinfo{year}{1973}).

\bibitem[{\citenamefont{Pal{\'a}sti}(1966)}]{palasti}
\bibinfo{author}{\bibfnamefont{I.}~\bibnamefont{Pal{\'a}sti}},
  \bibinfo{journal}{Studia Scient. Math. Hungar.} \textbf{\bibinfo{volume}{1}},
  \bibinfo{pages}{205} (\bibinfo{year}{1966}).

\bibitem[{\citenamefont{Karo{\'n}ski}(1982)}]{karonski}
\bibinfo{author}{\bibfnamefont{M.}~\bibnamefont{Karo{\'n}ski}},
  \bibinfo{journal}{J. Graph Theory} \textbf{\bibinfo{volume}{6}},
  \bibinfo{pages}{349} (\bibinfo{year}{1982}).

\bibitem[{\citenamefont{Ling and Killough}(1976)}]{killough}
\bibinfo{author}{\bibfnamefont{R.~F.} \bibnamefont{Ling}} \bibnamefont{and}
  \bibinfo{author}{\bibfnamefont{G.~G.} \bibnamefont{Killough}},
  \bibinfo{journal}{J. Amer. Statist. Assoc.} \textbf{\bibinfo{volume}{71}},
  \bibinfo{pages}{293} (\bibinfo{year}{1976}).

\bibitem[{\citenamefont{Dubes and Jain}(1979)}]{DJ}
\bibinfo{author}{\bibfnamefont{R.}~\bibnamefont{Dubes}} \bibnamefont{and}
  \bibinfo{author}{\bibfnamefont{A.~K.} \bibnamefont{Jain}},
  \bibinfo{journal}{Patt. Recog.} \textbf{\bibinfo{volume}{11}},
  \bibinfo{pages}{235} (\bibinfo{year}{1979}).

\bibitem[{\citenamefont{Slater}(1985{\natexlab{b}})}]{qq2}
\bibinfo{author}{\bibfnamefont{P.~B.} \bibnamefont{Slater}},
  \bibinfo{journal}{Quality and Quantity} \textbf{\bibinfo{volume}{19}},
  \bibinfo{pages}{211} (\bibinfo{year}{1985}{\natexlab{b}}).

\bibitem[{\citenamefont{Newman}(2004)}]{newman1}
\bibinfo{author}{\bibfnamefont{M.~E.~J.} \bibnamefont{Newman}},
  \bibinfo{journal}{Phys. Rev. E} \textbf{\bibinfo{volume}{70}},
  \bibinfo{pages}{056131} (\bibinfo{year}{2004}).

\bibitem[{\citenamefont{Nijenhuis and Wilf}(1975)}]{nijenhuis}
\bibinfo{author}{\bibfnamefont{A.}~\bibnamefont{Nijenhuis}} \bibnamefont{and}
  \bibinfo{author}{\bibfnamefont{H.~W.} \bibnamefont{Wilf}},
  \emph{\bibinfo{title}{Combinatorial Algorithms}}
  (\bibinfo{publisher}{Academic}, \bibinfo{address}{New York},
  \bibinfo{year}{1975}).

\bibitem[{\citenamefont{Slater}(1976{\natexlab{h}})}]{philippine}
\bibinfo{author}{\bibfnamefont{P.~B.} \bibnamefont{Slater}},
  \bibinfo{journal}{Philippine Geog. J.} \textbf{\bibinfo{volume}{20}},
  \bibinfo{pages}{79} (\bibinfo{year}{1976}{\natexlab{h}}).

\bibitem[{\citenamefont{Slater}(1977{\natexlab{b}})}]{brazil}
\bibinfo{author}{\bibfnamefont{P.~B.} \bibnamefont{Slater}},
  \bibinfo{journal}{Estad{\'i}stica} \textbf{\bibinfo{volume}{36}},
  \bibinfo{pages}{180} (\bibinfo{year}{1977}{\natexlab{b}}).

\bibitem[{\citenamefont{Slater}(1977{\natexlab{c}})}]{IO2}
\bibinfo{author}{\bibfnamefont{P.~B.} \bibnamefont{Slater}},
  \bibinfo{journal}{Empirical Econ.} \textbf{\bibinfo{volume}{3}},
  \bibinfo{pages}{49} (\bibinfo{year}{1977}{\natexlab{c}}).

\bibitem[{\citenamefont{Slater}(1979{\natexlab{b}})}]{cor}
\bibinfo{author}{\bibfnamefont{P.~B.} \bibnamefont{Slater}},
  \bibinfo{journal}{Comput. and Oper. Res.} \textbf{\bibinfo{volume}{6}},
  \bibinfo{pages}{205} (\bibinfo{year}{1979}{\natexlab{b}}).

\bibitem[{\citenamefont{Pentney and Meila}(20055)}]{pentney}
\bibinfo{author}{\bibfnamefont{W.}~\bibnamefont{Pentney}} \bibnamefont{and}
  \bibinfo{author}{\bibfnamefont{M.}~\bibnamefont{Meila}}, in
  \emph{\bibinfo{booktitle}{Proceedings of the National Conference on
  Artificial Intelligence}} (\bibinfo{year}{20055}), pp.
  \bibinfo{pages}{207--213}.

\bibitem[{\citenamefont{Dorogovtsev and Mendes}(2002)}]{evolution}
\bibinfo{author}{\bibfnamefont{S.~N.} \bibnamefont{Dorogovtsev}}
  \bibnamefont{and} \bibinfo{author}{\bibfnamefont{J.~F.~F.}
  \bibnamefont{Mendes}}, \bibinfo{journal}{Ann. Phys.}
  \textbf{\bibinfo{volume}{51}}, \bibinfo{pages}{1079} (\bibinfo{year}{2002}).

\end{thebibliography}

\end{document}